\documentclass[a4paper,fleqn]{cas-dc}
\usepackage{enumitem}
\usepackage[authoryear]{natbib}
\usepackage[figuresright]{rotating}
\usepackage{moreverb}
\usepackage{amsmath,amssymb,amsfonts}
\usepackage{algorithmic}
\usepackage{url}
\usepackage{graphicx}
\usepackage{textcomp}
\usepackage{xcolor}
\usepackage{tcolorbox}
\usepackage{ragged2e}
\usepackage{balance}
\usepackage{colortbl}
\usepackage{multirow}
\usepackage{color}
\usepackage{subcaption}
\usepackage{makecell}
\usepackage[linesnumbered, boxed,ruled]{algorithm2e}
\usepackage{CJKutf8}
\usepackage{siunitx}
\usepackage{rotfloat}
\usepackage{float}
\usepackage{lscape}
\usepackage{amssymb,mathrsfs,amsmath}
\usepackage{adjustbox}

\def\tsc#1{\csdef{#1}{\textsc{\lowercase{#1}}\xspace}}
\tsc{WGM}
\tsc{QE}
\tsc{EP}
\tsc{PMS}
\tsc{BEC}
\tsc{DE}
\newcommand{\fp}{\texttt{FasterPy}} 

\newif\ifcleanversion
\cleanversiontrue
\ifcleanversion
  \newcommand{\rev}[1]{#1}
  
  \newcommand{\todo}[1]{}
  
\else
  \newcommand{\rev}[1]{\textcolor{blue}{#1}}
  
  \newcommand{\todo}[1]{{\color{red}\textbf{[TODO: #1]}}}
  
\fi

\begin{document}
\begin{sloppypar}

\let\WriteBookmarks\relax
\def\floatpagepagefraction{1}
\def\textpagefraction{.001}
\shorttitle{FasterPy: An LLM-based Code Execution Efficiency Optimization Framework}
\shortauthors{Y Wu et al.}
\title [mode = title]{\texttt{FasterPy}: An LLM-based Code Execution Efficiency Optimization Framework}

\author[1]{Yue Wu}
\ead{iswuyue@whu.edu.cn}
\credit{Conceptualization, Investigation, Data curation, Formal analysis, Writing - Original draft preparation}

\author[1]{Minghao Han}
\ead{minghaoh@andrew.cmu.edu}
\credit{Conceptualization, Investigation, Data curation, Writing - Original draft preparation}

\author[1]{Peng Liang}
\cormark[1]
\ead{liangp@whu.edu.cn}
\credit{Conceptualization, Methodology, Investigation, Data curation, Supervision, Writing - Original draft preparation}
\address[1]{School of Computer Science, Wuhan University, China}

\author[1]{Ruiyin Li}
\cormark[1]
\ead{ryli_cs@whu.edu.cn}
\credit{Conceptualization, Methodology, Writing - review and editing}

\author[2]{Amjed Tahir}
\ead{a.tahir@massey.ac.nz}
\credit{Conceptualization, Methodology, Writing - review and editing}
\address[2]{School of Mathematical and Computational Sciences, Massey University, New Zealand}

\author[3]{Zengyang Li}
\ead{zengyangli@ccnu.edu.cn}
\credit{Conceptualization, Methodology, Writing - review and editing}
\address[3]{School of Computer Science \& Hubei Provincial Key Laboratory of Artificial Intelligence and Smart Learning, Central China Normal University, China}

\author[4]{Qiong Feng}
\ead{qiongfeng@njust.edu.cn}
\credit{Conceptualization, Methodology, Writing - review and editing}
\address[4]{School of Computer Science, Nanjing University of Science and Technology, China}

\author[5]{Mojtaba Shahin}
\ead{mojtaba.shahin@rmit.edu.au}
\credit{Conceptualization, Methodology, Writing - review and editing}
\address[5]{School of Computing Technologies, RMIT University, Australia}

\cortext[cor1]{Corresponding author.}


\begin{abstract}
Code often suffers from performance bugs. These bugs necessitate the research and practice of code optimization. Traditional rule-based methods rely on manually designing and maintaining rules for specific performance bugs (e.g., redundant loops, repeated computations), making them labor-intensive and limited in applicability. In recent years, machine learning and deep learning-based methods have emerged as promising alternatives by learning optimization heuristics from annotated code corpora and performance measurements. However, these approaches usually depend on specific program representations and meticulously crafted training datasets, making them costly to develop and difficult to scale. With the booming of Large Language Models (LLMs), their remarkable capabilities in code generation have opened new avenues for automated code optimization. In this work, we proposed \fp{}, a low-cost and efficient framework that adapts LLMs to optimize the execution efficiency of Python code. \fp{} combines Retrieval-Augmented Generation (RAG), supported by a knowledge base constructed from existing performance-improving code pairs and corresponding performance measurements, with Low-Rank Adaptation (LoRA) to enhance code optimization performance. \rev{Our experimental results on the \textbf{P}erformance-\textbf{I}mproving Code \textbf{E}dits (PIE) benchmark demonstrate the effectiveness of our method across multiple metrics.} The \fp{} tool and the experimental results are available at \url{https://github.com/WuYue22/fasterpy}.
\end{abstract}


\maketitle

\section{Introduction}\label{Introduction}
In recent years, the problem of improving code execution efficiency has received sustained attention in the field of software engineering~\citep{Giavrimis2025Artemis, Chen2023SupersonicLT, Madaan2023LearningPC, Rahman2025MARCOAM, Ye2025LLM4EFFILL, Bai2025POLO}. Traditional rule-based methods mainly focus on optimizing specific performance bugs, such as redundant loops and repeated computations, and have achieved promising results~\citep{Toffola2015Pb, Olivo2015CLARITY}. However, designing such rules is often costly, and these methods lack adaptability to unseen scenarios~\citep{Toffola2015Pb, Olivo2015CLARITY, Wang2018Machine}. More recently, new algorithms have been introduced to support code optimization~\citep{Giavrimis2021Genetic, Lopez2018Source}, but their performance often depends heavily on specific code structures and the definitions of target functions to be optimized. To overcome these limitations, researchers have explored machine learning and deep learning methods for code optimization~\citep{Chris2017E2EDL, Lamouri2025Pearl,Bendib2024ARL}. These methods typically learn optimization heuristics from annotated code corpora and performance measurements, showing the potential to capture patterns beyond manually designed rules. Nevertheless, their effectiveness often depends heavily on specific program representations (e.g., LLVM Intermediate Representation, Abstract Syntax Trees) and carefully crafted training datasets, which are costly and limit the generalizability of these methods across diverse programming tasks~\citep{Ashouri2018SurveyCA}. With the rapid advancement of LLMs~\citep{chatgpt, Shao2024DeepSeekV2AS, claude, gemini, llama}, which have profoundly influenced the field of software engineering, LLMs have demonstrated remarkable capabilities in code understanding and generation tasks, leading to their widespread adoption in software development and paving the way for LLM-based code optimization~\citep{Yi2025AnES}. 
In this context, code efficiency is particularly critical in both research and real-world applications, as the computational efficiency of programs is a fundamental cornerstone of system performance and user satisfaction~\citep{iso25040}. 

In this study, we focus specifically on the execution efficiency of code, which is primarily measured by execution time~\citep{Gong2025TRACY}. To improve the execution efficiency of code and address the limitations of existing approaches, we propose \textbf{\fp{}}, a low-cost and high-efficiency framework for optimizing code execution performance. Our work targets function-level optimization, aiming to improve the runtime performance of Python code without altering their semantics. We consider Python code as the target language in this study, as it is widely used in practice, especially among AI engineers, and its runtime overhead makes it a natural candidate for optimization~\citep{Stoico2025AnES}. By combining Retrieval-Augmented Generation (RAG) with Low-Rank Adaptation (LoRA) fine-tuning, the \fp{} framework adaptively enhances LLMs for code efficiency optimization tasks. It consists of two main components:
\textbf{(1)} A \textit{Code Efficiency Optimization Knowledge Base}, which stores low-efficiency (slow) code snippets along with their corresponding optimization suggestions. This component enables semantic similarity matching based on the input code and provides relevant suggestions as references for code improvement.
\textbf{(2)} A \textit{Fine-Tuned Code Generation Model}, obtained by applying LoRA fine-tuning to an existing LLM, which automatically generates an optimized version of the code based on the input code and the optimization suggestions retrieved in (1). These two components form a feedback loop system from semantic retrieval to efficiency-oriented code generation.

Our main \textbf{contributions} are twofold: 
\textbf{(1)} \textit{A Code Efficiency Optimization Knowledge Base.} We constructed a retrieval-oriented knowledge base grounded in the Mercury~\citep{du2024mercury} and \textbf{P}erformance-\textbf{I}mproving Code \textbf{E}dits (PIE)~\citep{Madaan2023LearningPC} benchmarks. Beyond integrating optimization instances from the two datasets, we augment them with LLM-generated summaries that capture the underlying optimization strategy and code transformation intent. We further reorganize the knowledge base into a unified retrieval-oriented structure, enabling effective retrieval of relevant optimization patterns and providing diverse optimization cases to support retrieval-augmented LLM-based code optimization.
\textbf{(2)} \textit{A Code Execution Efficiency Optimization Framework.} We proposed \fp{}, a cost-effective and efficient framework for optimizing code execution performance using LLMs.  
Designed with practicality in mind, \fp{} supports code optimization under standard hardware settings, without relying on specialized or high-cost computational resources. The framework's components, including the optimization knowledge base and the fine-tuned model, are modular and extensible, making it potentially adaptable to other programming languages and promising for broader adoption. Preliminary results on C++ provide initial evidence of this potential.


\textbf{Organization}: The rest of this paper is structured as follows: Section~\ref{Background} presents the relevant background and covers the main concepts used in this study. Section \ref{Related Work} presents the related work. Section \ref{Approach} provides a detailed description of \fp{}, including its core modules and implementation methods. Section~\ref{Experimental Design} describes the experimental design, including the data processing pipeline, the experimental setup, and the improvements made to the PIE benchmarking program. Section~\ref{Evaluation} presents the evaluation methodology, benchmarks, metrics, and results. Section~\ref{Ablation Study} presents the ablation studies, including component ablation, retrieved context representation ablation, and dataset ablation. The findings are further analyzed and discussed in Section~\ref{Discussion}. The potential threats to validity are clarified in Section \ref{Threats to Validity}. Finally, Section \ref{Conclusions} concludes this work with future work directions.

\section{Background} \label{Background}
This section provides the necessary background for our work, focusing on two key techniques: Retrieval-Augmented Generation (RAG) and Low-Rank Adaptation (LoRA). 

\rev{RAG integrates LLMs with external information retrieval systems, allowing relevant knowledge to be retrieved and incorporated into the generation process~\citep{Lewis2020RAGKIT}. By providing task-relevant external knowledge together with the original input, RAG can improve model performance on knowledge-intensive tasks~\citep{Gao2023retrieval, Fan2024SurveyRag}. In this work, we adopt the RAG technique to provide LLMs with external code optimization knowledge. Specifically, the input code is transformed into a vector representation and used to retrieve semantically relevant code samples and optimization suggestions from a knowledge base. The retrieved suggestions are then provided to the LLM as additional context for code optimization.}



\rev{LoRA is a parameter-efficient fine-tuning (PEFT) method that adapts pretrained models to downstream tasks by introducing trainable low-rank matrices while keeping the original model parameters frozen~\citep{Hu2022LoRA, Han2024PEFTS}. Compared with full-parameter fine-tuning, LoRA substantially reduces the number of trainable parameters and associated computational resources. In this work, we use LoRA to adapt pretrained LLMs to the code optimization task, enabling them to better follow the input-output format and optimization objectives of the task.}

\section{Related Work} \label{Related Work}
In this section, we present the related work in two aspects, i.e., traditional approaches to code optimization (Section~\ref{traditional approaches to code optimization}) and LLMs for code optimization (Section~\ref{llms for code optimization}).

\subsection{Traditional Approaches to Code Optimization}\label{traditional approaches to code optimization} 
Early studies on code optimization primarily focused on rule-based methods, which leverage expert knowledge bases to optimize specific types of performance bugs and have proven effective in practice. ~\cite{Toffola2015Pb} introduced MemoizeIt, which employs dynamic analysis to compare the inputs and outputs of method calls, thereby helping developers identify cases of repeatedly executed computations. MemoizeIt was evaluated on 11 real-world Java programs, achieving speedups ranging from 1.04× to 12.93×. Complementarily, ~\cite{Olivo2015CLARITY} proposed CLARITY, a static analysis approach that automatically detects redundant traversal bugs. CLARITY was demonstrated on nine open-source Java code bases and achieved performance improvements of at least 2.45× on large inputs.

Optimization algorithms were also introduced into the field of code optimization. ~\cite{Lopez2018Source} enhanced code performance through the replacement of equivalent mutation operators. ~\cite{Giavrimis2021Genetic} employed genetic algorithms to select implementations of abstract data structures in C++ code to improve performance. It achieved up to 27.9\% faster runtime, 16.1\% lower CPU usage, and 2.7\% lower memory usage in preliminary tests on three mainstream C++ libraries. ~\cite{liu2024efficient} significantly reduced the search space of the phase-ordering problem by modeling source code and performance dependencies between compiler optimization passes and using a clustering algorithm to automatically group these passes into efficient subsequences, ultimately achieving a 22\% improvement in runtime performance and a 24\% reduction in code size.

Recent years have witnessed the emergence of Machine Learning (ML) and Deep Learning (DL)-based methods demonstrating significant potential in code optimization. ~\cite{Deng2024CompilerDreamLA} introduced CompilerDream, a model-based reinforcement learning framework for general code optimization, which learns a ``compiler world model'' to efficiently train an optimization policy that achieves strong zero-shot generalization and outperforms LLVM's built-in optimizations on unseen programs. ~\cite{Bendib2024ARL} developed a reinforcement learning environment for the Multi-Level Intermediate Representation (MLIR) compiler, where they trained an agent to automatically optimize code that results in execution time faster than the optimized kernels generated by TensorFlow Just-In-Time compiler. ~\cite{Lamouri2025Pearl} introduced Pearl, a deep reinforcement learning framework that learns general code optimization strategies capable of generalizing to unseen programs, significantly outperforming top-tier compilers like Tiramisu by a geometric mean speedup of 2.02x.

\subsection{LLMs for Code Optimization}\label{llms for code optimization}
With the rapid advancement of LLMs, the success of Transformer-based code models such as Codex~\citep{Chen2021Codex} has opened up new possibilities for automated optimization of code execution efficiency. ~\cite{Garg2022DeepPERFAD} proposed DeepPERF, a Transformer-based model that is pretrained on mixed English-code data and fine-tuned using commit histories from GitHub repositories. DeepPERF generates performance-enhancing patches for C\# programs. ~\cite{Florath2023LLMIO} utilized interactive human-LLM collaboration with GPT-4 to optimize critical functions in the open-source Python libraries Pillow~\citep{pillow} and Numpy~\citep{numpy}, achieving performance speedups of up to 38x. ~\cite{Chen2023SupersonicLT} proposed SUPERSONIC, a sequence-to-sequence model fine-tuned from CodeBERT that achieves precise optimization of C/C++ source code by learning to generate patches in a diff format, with its performance on this specific task significantly surpassing GPT-3.5 and GPT-4. ~\cite{Madaan2023LearningPC} explored multiple strategies to adapt LLMs for code optimization tasks, including prompt engineering, RAG, and fine-tuning, leveraging edit histories from competitive programming datasets. ~\cite{Gao2024SearchBasedLF} proposed a search-based code optimization framework for Python and C++ named SBLLM, which integrates LLMs with evolutionary search via adaptive pattern retrieval and genetic operator-inspired prompting, achieving program execution efficiency improvements of up to 209.59\% and consistently outperforming baseline methods by 8.75\%–28.06\% on Python and 1.15\%–9.56\% on C++, demonstrating its effectiveness in iteratively discovering optimal code optimizations. ~\cite{Rahman2025MARCOAM} proposed Multi-Agent Reactive Code Optimizer (MARCO), which leverages a specialized agent architecture and an adaptive feedback mechanism to optimize high-performance computing (HPC) code. Compared to Claude 3.5 Sonnet, MARCO achieved a 14.6\% average runtime reduction. ~\cite{Giavrimis2025Artemis} proposed the Artemis AI framework, which collaboratively leverages multiple LLMs and structures the code optimization process into a systematic, search-based, multi-stage workflow. Their approach seeks to find solutions that achieve maximum performance impact with minimal code changes. Artemis AI demonstrated concrete performance improvements on several real-world projects, achieving execution time reductions ranging from 15\% to 52\%. To overcome the limitation of Artemis where prompts optimized for one model perform poorly on another, ~\cite{Gong2025TuningLC} proposed a framework named Meta-Prompted Code Optimization (MPCO). This framework enhances LLM-based code optimization by automatically generating prompts tailored to specific tasks and model characteristics for diverse LLMs, achieving performance improvements of up to 19.06\% across five real-world codebases. ~\cite{Zhao2025SemOptLC} proposed a code optimization framework for C/C++ named SemOpt, which combines LLMs with rule-based static analysis. SemOpt achieved performance improvements on test cases in real-world projects ranging from 5.04\% to 218.07\%, and manual evaluation verified that 89.86\% of its generated optimization suggestions improved performance while preserving semantic correctness, demonstrating the practical utility of the method.

\subsection{Conclusive Summary}\label{conclusive summary} 
While the aforementioned approaches have demonstrated promising results in code execution efficiency optimization, several limitations remain. First, rule-based code optimization methods demonstrate the effectiveness of expert-crafted rule libraries in addressing specific types of inefficiencies; however, they suffer from limited generalizability and scalability, and constructing and maintaining these knowledge bases incurs high costs. Second, algorithmic optimization approaches often rely heavily on specific code structures and the precise definition of objective functions, which restricts their applicability across diverse codebases. Third, ML and DL-based methods usually depend on specific program representations and meticulously crafted training datasets, making them costly to develop and difficult to scale. Finally, research on LLM-based code execution efficiency optimization also faces two challenges: (i) many models are tailored to specific programming languages or benchmark settings, limiting their flexibility. For example, DeepPERF~\citep{Garg2022DeepPERFAD} is designed for only C\# programs; and (ii) some approaches suffer from suboptimal optimization efficiency or incur high computational costs, making them difficult to deploy in resource-constrained environments. For example, MARCO~\citep{Rahman2025MARCOAM} incurs computational overhead through iterative communication between agents and the transmission of intermediate results. These limitations highlight the need for a more flexible and efficient optimization framework.

The components of our proposed \fp{} framework (e.g., the knowledge base and fine-tuned models) are designed to be replaceable and extensible, ensuring strong practicality and flexibility. Moreover, the framework enables efficient optimization without relying on costly hardware resources.

\section{Approach} \label{Approach}
In this section, we provide a detailed description of our proposed approach \fp{}. We begin with an overview of the entire framework in Section~\ref{fasterpy Overview}, outlining its workflow and key modules. The following sections provide detailed descriptions of each step, including the code preprocessing step in Section~\ref{Code Preprocessing}, the code embedding step in Section~\ref{Code Embedding}, the knowledge base retrieval step in Section~\ref{Knowledge Base Retrieval}, and the code generation step in Section~\ref{Code Generation}.

\subsection{Overview of \fp{}}\label{fasterpy Overview}
\begin{figure*}[htbp]
    \centering
    \includegraphics[width=\linewidth]{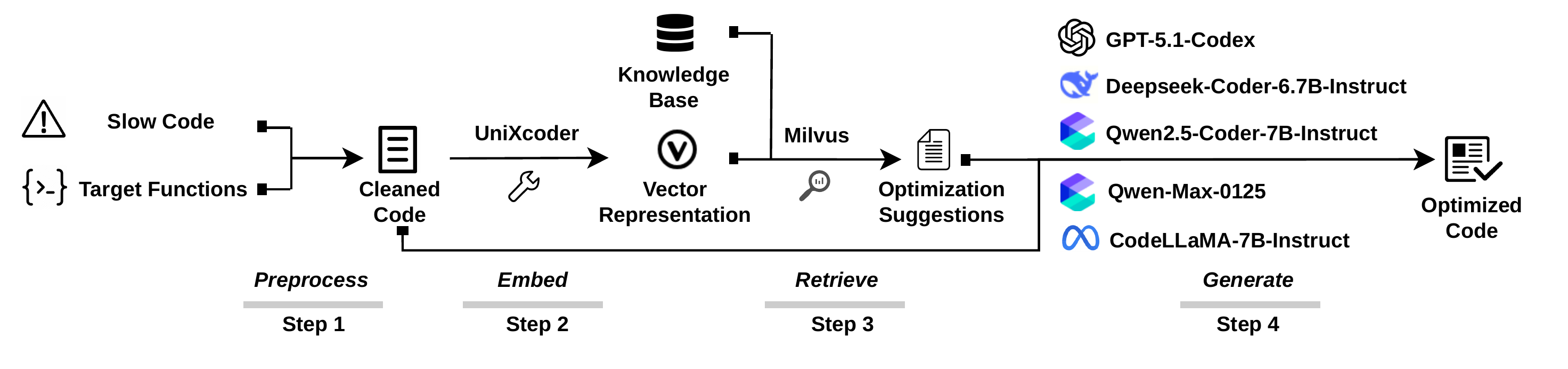}
    \caption{Overview of the \fp{} Framework}
    \label{fig:overview}
\end{figure*}

\fp{} integrates a code optimization knowledge base and a LoRA-fine-tuned LLM, which together enable an end-to-end workflow: retrieving optimization guidance from the knowledge base and applying the guidance with the fine-tuned model to generate functionally equivalent code with improved execution efficiency. As shown in Figure~\ref{fig:overview}, our framework begins with the user's input, which includes a source code file containing inefficient implementations (i.e., \textit{slow code}) and a set of functions to be optimized, identified by their names (i.e., \textit{target functions}). In this work, we focus on improving the execution efficiency of Python code. The scope of inefficiency we target includes common inefficiency patterns such as suboptimal algorithmic complexity (e.g., brute-force algorithms), inefficient use of data structures and memory access patterns, unnecessary library imports, manual implementations of operations that could be efficiently performed using built-in functions, and the replacement of high-level optimized constructs (e.g., vectorized operations) with slower explicit loops. The overall workflow of the \fp{} framework consists of the following four steps, each of which is implemented by a dedicated and replaceable module.

\textbf{Step 1. Code Preprocessing.}
In the first step, the framework extracts the target functions and their relevant contextual information from the original code (denoted as $C_{origin}$), including function bodies, global variables, class definitions, and other invoked functions. Irrelevant function declarations and definitions are removed, resulting in a cleaned code snippet referred to as $C_{cleaned}$.

\textbf{Step 2. Code Embedding.}
In this step, the cleaned code snippet $C_{cleaned}$ is converted into a semantic representation vector $E_{cleaned}$ through the code embedding process.

\textbf{Step 3. Knowledge Base Retrieval.}
The vector $E_{cleaned}$ is then fed into the knowledge base retrieval module, which retrieves the top $m$ semantically similar \textit{slow code} snippets from a preconstructed optimization knowledge base based on vector similarity. For each retrieved snippet, this step also obtains its associated $n$ optimization suggestions along with their corresponding performance improvement measurements. It then selects the top $k$ suggestions that balance semantic relevance and expected optimization benefit, and aggregates them into a suggestion set denoted as $R_k$.

\textbf{Step 4. Code Generation.}
After retrieving the candidate $k$ suggestions, the framework integrates $C_{cleaned}$ with the suggestion set $R_k$ to construct a unified input prompt denoted as $P$, which incorporates both the target code and the optimization guidance. The prompt $P$ is then fed into a fine-tuned LLM to generate an optimized version of the code. The output $C_{optimized}$ marks the end of the optimization pipeline.

\subsection{Code Preprocessing}\label{Code Preprocessing}
In practice, source code often exists within a complex contextual environment. The execution behavior of a function depends not only on its internal structure, but also on external factors such as global variables, class members, and calls to other functions. Consequently, directly providing only the target function as input to the LLM may fail to capture its actual execution semantics, while using the entire source file may introduce excessive irrelevant information. Such noisy or unrelated content can negatively affect both semantic understanding and subsequent optimization quality. Since the effectiveness of our framework ultimately depends on the quality of the input provided to knowledge base retrieval and code generation, an appropriate preprocessing strategy is necessary to preserve execution-relevant information while reducing unnecessary complexity.

Our preprocessing step follows the intuition of human engineers: it retains the body of the target function while preserving as much of its essential context as possible, such as related variables and invoked functions. Meanwhile, Cho et al. have shown that the performance of LLMs tends to degrade as the input sequence length increases~\citep{Cho2014OnTP}. In addition, excessively long or redundant inputs may also trigger hallucinations in LLMs~\citep{Barkley2024InvestigatingTR}, leading to unrealistic or unreasonable optimization outputs. Since our framework ultimately relies on an LLM for code optimization, it is important to limit the input length while retaining essential context. Therefore, beyond preserving the necessary contextual information, one of the key responsibilities of the preprocessing step is to control the complexity and length of the input code.

\begin{figure*}[htbp]   
  \centering            
  \subfloat[The original code to be preprocessed]   
  {
      \includegraphics[width=0.4\textwidth]{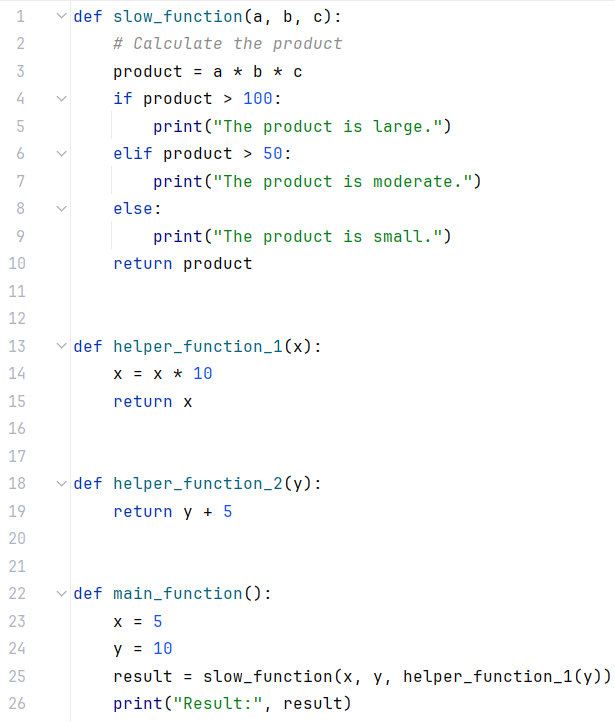}
      \label{fig: original source code}
  }
  \hfill
  \subfloat[AST after Step 1 (Syntactic Parsing)]
  {
      \label{fig: AST after step 1}\includegraphics[width=0.4\textwidth]{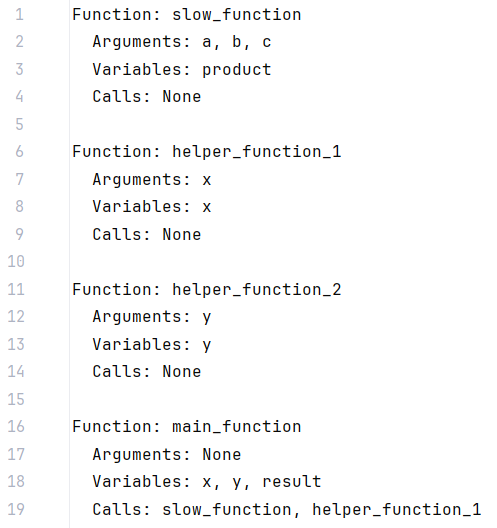}
  }
  \hfill
  \subfloat[Code after Steps 2 and 3 (Context Extraction and AST-Guided Cleaning)]
  {
      \label{fig: code after steps 2 and 3}\includegraphics[width=0.4\textwidth]{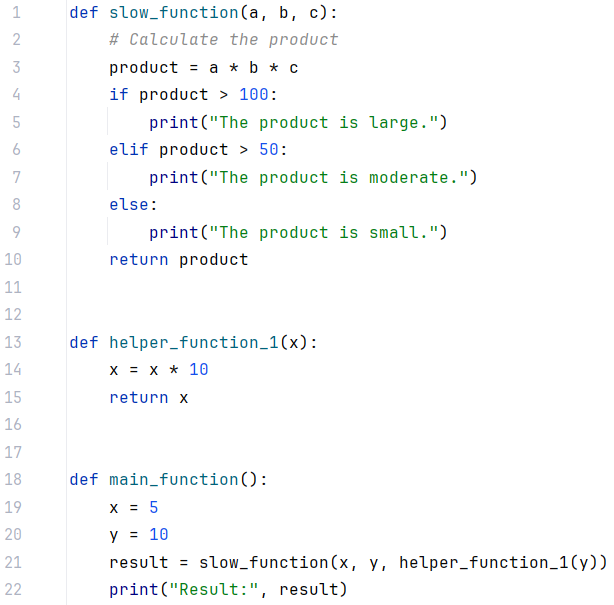}
  }
  \hfill
  \subfloat[Code after Step 4 (Code Formatting)]
  {
      \label{fig: code after step 4}\includegraphics[width=0.4\textwidth]{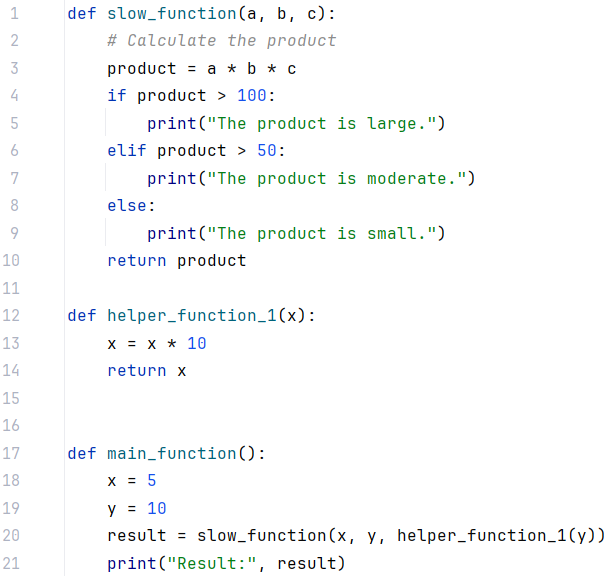}
  }
  \caption{Example of Code Preprocessing} 
  \label{fig: example of code preprocessing}  
\end{figure*}

Based on the above considerations, our preprocessing step proceeds as follows. To illustrate the process, we provide an example of how the code evolves through each step. The original source code is shown in Figure~\ref{fig: original source code}.

\textbf{Step 1.1. Syntactic Parsing}: We use Tree-sitter with its Python grammar, tree-sitter-python\footnote{\url{https://github.com/tree-sitter/py-tree-sitter}}, to perform syntactic parsing on the original source code and construct its Abstract Syntax Tree (AST). As shown in Figure~\ref{fig: AST after step 1}, from the AST, we can clearly identify the functions, variables, and their dependencies.

\textbf{Step 1.2. Context Extraction}: Based on the set of target function names, we retain the body of each function along with its relevant variable declarations, class definitions, and all invoked functions, which are identified through the analysis of these functions. As shown in Figure~\ref{fig: code after steps 2 and 3}, we extracted the target function \texttt{slow\_function} along with its relevant context, retaining the variables \texttt{a}, \texttt{b}, and \texttt{c}, as well as the function \texttt{helper\_function\_1} that are related to its execution. Additionally, we keep the relevant content in \texttt{main\_function}.

\textbf{Step 1.3. AST-Guided Cleaning}: Using line number information from the AST, we build a unified-style diff patch to remove unrelated functions. As shown in Figure~\ref{fig: code after steps 2 and 3}, we remove the irrelevant content from the target function, such as deleting \texttt{helper\_function\_2} since it is not called.

\textbf{Step 1.4. Code Formatting}: We apply the Black formatter~\citep{black} to the retained code, producing a well-structured and human-readable standardized input text. As shown in Figure~\ref{fig: code after step 4}, after applying the Black formatter, the indentation and spacing are in accordance with the standard, and unnecessary blank lines are removed.

The preprocessing step not only reduces the length of the input code and helps the LLM process the code more effectively, but also preserves as much semantic information as possible that is relevant to the target functions. This sets a solid foundation for the efficient operation of the subsequent embedding and optimization steps. The complete preprocessing step is detailed in Algorithm~\ref{Code Preprocessing Procedure Algorithm}.

\begin{algorithm}[htbp]
  \SetAlgoLined
  \small
  \SetKwInOut{Input}{input}\SetKwInOut{Output}{output}

  \Input{ origin\_code, target\_functions}
  \Output{ cleaned\_code}
  \If{target\_functions == None}{return origin\_code}
  $syntax\_tree\leftarrow parse(origin\_code)$\;
  $preserve\_funcs\leftarrow set(target\_functions)$\;
  \ForEach{preserve\_func in preserve\_funcs}
  {
    $callees\leftarrow recursion\_get\_callees(preserve\_func,{})$\;
    preserve\_funcs.update(callees);
  }
  $delete\_hunks\leftarrow [ ]$\;
  $all\_function\_nodes\leftarrow get\_all\_function\_nodes(syntax\_tree)$\;
  \ForEach{function\_node in all\_function\_nodes}
  {
  \eIf{$function\_node.name \in preserve\_funcs$}{continue;}{
    delete\_hunks.append(function\_node.start\_line, function\_node.end\_line);
  }
  }
  $delete\_hunks\leftarrow sort\_by\_start\_line(delete\_hunks)$\;
  $merged\leftarrow delete\_hunks[0] $\;
  \ForEach{(start,end) in delete\_hunks[1:]}
  {
    $(last\_start,last\_end)\leftarrow merged[-1]$\;
    \eIf{$start \leq last\_end$}{$merged[-1]\leftarrow (last\_start, max(last\_end,end)) $\;}{
    $merged.append(start,end)$;
  }
  }
  $delete\_hunks\leftarrow merged$\;
  $patch\leftarrow generate\_patch(origin\_code, delete\_hunks)$\;
  $cleaned\_code\leftarrow apply\_patch(origin\_code, patch)$\;
  $ cleaned\_code\leftarrow format(cleaned\_code)$\;
  \Return cleaned\_code;
  
  \caption{Code Preprocessing Procedure}
  \label{Code Preprocessing Procedure Algorithm}
\end{algorithm}

\subsection{Code Embedding}\label{Code Embedding}
The effectiveness of RAG largely depends on the ability to accurately retrieve highly relevant information from the knowledge base. In \fp{}, the core of this mechanism lies in whether the system can, based on the cleaned code snippet $C_{cleaned}$, successfully retrieve the $m$ most semantically similar \textit{slow code} snippets and obtain their associated optimization suggestions and performance improvement measurements. Only when semantic alignment is ensured during the retrieval step can the subsequent code generation step be grounded in meaningful and effective references.

To achieve the above objective, this work adopts vector-based retrieval in place of traditional string matching. Although string-based exact matching can be efficient in certain scenarios, it often fails to capture code snippets that are semantically similar but differ in the syntax level, such as variable or function names~\citep{Cheng2022CSRSCS}. This limitation can significantly degrade the quality of knowledge base retrieval. Therefore, we represent code in a distributed vector space and use vector similarity as the retrieval metric, enabling stronger semantic generalization during the retrieval process.

We adopt the Encoder-Only mode of the UniXcoder~\citep{Guo2022UniXcoderUC} (a multimodal code representation model) to obtain the vector representation of the cleaned code $C_{cleaned}$, denoted as $E_{code}$. The $E_{code}$ is then normalized using L2 normalization~\citep{Salton1975VSM}, as shown in Equation~\ref{l2}, to facilitate subsequent calculations of the inner product. The normalized embedding is denoted as $E_{cleaned}$. Compared to purely text-based encoders, UniXcoder offers two major advantages: (1) Multimodal input: During encoding, UniXcoder incorporates not only the raw code text but also the comment information and the structure of the AST. This enables the model to capture syntactic, semantic, and documentation-level signals simultaneously during embedding, thereby improving the completeness and accuracy of code representations. (2) Sentence-level representation with fixed dimensionality: In the output stage, UniXcoder applies mean pooling over token-level embeddings, producing a fixed-dimensional vector representation for code inputs of varying lengths.
\begin{equation}
    \hat{\mathbf{E}}=\frac{\mathbf{E}}{\|\mathbf{E}\|^{2}}=\frac{\mathbf{E}}{\sqrt{\sum_{i=1}^{d} E_{i}^{2}}}
    \label{l2}
\end{equation}

\subsection{Knowledge Base Retrieval}\label{Knowledge Base Retrieval}
Although LLMs have demonstrated strong capabilities in code-related tasks, relying solely on model parameters may not always provide sufficient guidance for code performance optimization. Therefore, our framework retrieves relevant optimization examples together with their associated optimization suggestions and incorporates them as external knowledge to guide the subsequent optimization process. The core task of the knowledge base retrieval step is to efficiently retrieve the top-$m$ records from the knowledge base whose \textit{slow code} snippets are semantically most similar to the normalized vector representation $E_{cleaned}$, which is produced by the code embedding module. For each retrieved record, the corresponding optimization suggestions and historical performance improvement measurements are also obtained. This retrieval process provides actionable references to build the subsequent optimization prompt and directly influences the final optimization quality.

We adopt Milvus\footnote{\url{https://milvus.io}}, a vector database, as a retrieval engine to build a knowledge base for code execution efficiency optimization. This database supports high-dimensional vector search and provides Approximate Nearest Neighbor (ANN) capabilities, making it well-suited for large-scale semantic retrieval tasks. The schema of the knowledge base is shown in Table~\ref{tab:optimization_schema}. The construction of the knowledge base and its data sources will be detailed in Section~\ref{Knowledge Base Construction}.

\begin{table}[h]
    \centering
    \caption{Schema of the Optimization Suggestion Knowledge Base}
    \begin{tabular}{>{\raggedright\arraybackslash}p{0.9cm}>{\raggedright\arraybackslash}p{4.7cm}>{\raggedright\arraybackslash}p{1.2cm}}
    \toprule 
        \textbf{Name} & \textbf{Description} & \textbf{Type} \\
    \midrule 
        id & Unique identifier of the optimization case & INT64 \\
        
        
        \hline vector & Semantic vector representation of the slow code & \makecell[l]{FLOAT\_\\VECTOR} \\
        
        \hline summary & Optimization suggestions for improving the execution efficiency of the slow code & VARCHAR \\
        
        \hline rate & Performance improvement score of the suggestion in historical experiments & FLOAT \\
        \bottomrule 
    \end{tabular}
    \label{tab:optimization_schema}
\end{table}

During the retrieval step, the input vector $E_{cleaned}$ is used as the query vector. Milvus employs inner product similarity as the evaluation metric to compute the similarity between the query vector and the vector representations of \textit{slow code} snippets stored in each record in the database - measuring how close the input code $C_{cleaned}$ is to each stored \textit{slow code} in the semantic space. The system returns the top-$m$ records with the highest similarity scores above a predefined absolute similarity threshold $\theta$, denoted as the set $R_m$, in which each record contains its full metadata fields and the associated similarity score.

Since the knowledge base is primarily constructed from online sources, as detailed in Section~\ref{Knowledge Base Construction}, the effectiveness of the associated optimization suggestions can vary significantly. Although the previous retrieval step ensures the semantic relevance of the suggestions, relying solely on semantic similarity for ranking may fail to fully utilize suggestions that are slightly less similar but offer better optimization performance. For example, some \textit{slow code} snippets with high semantic similarity to the input may correspond to suggestions that perform poorly in practice, while others with lower similarity scores may provide more effective optimizations. To improve the practical utility of the retrieved suggestions, we apply a filtration process to the initial candidate set $R_m$ as detailed in Algorithm~\ref{Suggestions Filtering Algorithm}. This filtering process considers both semantic similarity and optimization effectiveness as joint criteria.

\begin{algorithm}[htbp]
  \SetAlgoLined
  \SetKwInOut{Input}{input}\SetKwInOut{Output}{output}

  \Input{$R_m$, top\_k, distance\_range}
  \Output{suggestions}
  \If{$ \left|R_m\right| \leq top\_k$}{
    suggestions.append($R_m$)\;
    \Return suggestions;}
  $sort\_by\_distance\leftarrow desc\_sort(R_m, key = distance)$\;
  \rev{$suggestions\leftarrow [sort\_by\_distance[0]]$\;
  $minimum\_distance\leftarrow sort\_by\_distance[0]['distance']-distance\_range$\;
  $sort\_by\_distance\leftarrow sort\_by\_distance[1:]$\;}
  $sort\_by\_distance\leftarrow$ $[R_m$ for $R_m$ in $sort\_by\_distance$ if $R_m['distance'] \geqslant$ $minimum\_distance]$\;
  \If{$\left|sort\_by\_distance\right| < top\_k -1 $}{
  suggestions.extend(sort\_by\_distance)\;
  \Return suggestions;
  }
  $sort\_by\_rate\leftarrow desc\_sort(R_m, key=rate)$\;
  suggestions.extend(sort\_by\_rate[:top\_k-1])\;
  \Return suggestions;
  
  \caption{Suggestions Filtering}
  \label{Suggestions Filtering Algorithm}
\end{algorithm}

The filtration process proceeds as follows:

\textbf{Step 3.1.} Retain the most semantically similar record $R_{closest}$ from the candidate set $R_m$, with its similarity to $C_{cleaned}$ denoted as $S_{closest}$.

\textbf{Step 3.2.} Filter the remaining records by keeping only those that satisfy the condition $S_i \geq S_{\text{closest}} - \Delta_s$, where $\Delta_s$ is a predefined similarity tolerance threshold.

\textbf{Step 3.3.} Sort the remaining candidates in descending order according to their performance improvement scores (denoted as $rate$), and select the top $k-1$ suggestions.

\textbf{Step 3.4.} Return the final set of $k$ suggestions by combining the $k-1$ selected suggestions with $R_{closest}$.

This dual-criteria filtration mechanism not only ensures the relevance of the recommended suggestions through a similarity threshold, but also guaranties their practical effectiveness by selecting high-$rate$ candidates. It enhances the ability of the subsequent generation module to produce efficient code while reducing the risk of hallucinations or performance degradation caused by inappropriate suggestions.

\subsection{Code Generation}\label{Code Generation}
The primary task of the optimization module is to automatically generate an optimized version of the code, denoted as $C_{optimized}$, with improved execution efficiency. This is achieved based on the cleaned code snippet $C_{cleaned}$ produced by preprocessing module and the top-$k$ high-quality optimization suggestions $R_k$ retrieved from the knowledge base. This process can be divided into three steps: prompt construction, code generation, and post-processing.

\textbf{Step 4.1. Prompt Construction.} The cleaned input code $C_{cleaned}$ is first combined with the top-$k$ retrieved suggestions $R_k$ to form a unified prompt input. The prompt template is shown below. Specifically, the prompt template adopts a role-playing instruction to place the LLM in a code optimization scenario, explicitly defines the objective of reducing runtime while maintaining correctness, and provides retrieved optimization suggestions as optional guidance rather than strict constraints.

\begin{tcolorbox}[colback=gray!10, colframe=gray!80, title=Prompt Template for Code Optimization]
\textbf{System Prompt:}

You are an AI programming assistant. Try your best to optimize the Input code by focusing on reducing its runtime while maintaining correctness. Output the optimized code. Feel free to refer to the suggestions below for potential optimizations, but you’re not restricted to them. Applicability represents the degree to which a suggestion fits the input code. Rate represents the degree to which a suggestion improves the input code.
\vspace{0.2cm}

\textbf{User Prompt:}

Input code:

\verb|```|

 $C_{cleaned}$
 
\verb|```|

\vspace{0.2cm}
Suggestions:

Applicability: $R_i$.distance; Rate: $R_i$.rate; $R_i$.text

\vspace{0.2cm}
Output:

\verb|```|
 
$C_{optimized}$
 
\verb|```|
 
\label{prompt: template}
\end{tcolorbox}

It is worth noting that although each record in $R_k$ typically contains the original slow code snippet, its corresponding optimization suggestion, and a performance score (i.e., $rate$), we deliberately exclude the slow code content during prompt construction and retain only the optimization suggestions. This design is based on our empirical findings: when using LLMs with fewer than 10 billion parameters, including both $C_{cleaned}$ and multiple slow code examples in the prompt, the model can be easily misled. This often results in the generated $C_{optimized}$ being contaminated with unrelated variable names, function structures, or execution paths from the slow code, leading to semantic inconsistencies or hallucinations. In contrast, providing only natural language suggestions helps the model focus more effectively on the semantic structure and performance bottlenecks of $C_{cleaned}$, producing more reliable optimization results.

\textbf{Step 4.2. Code Generation.} After the prompt is constructed, it is fed into an LLM that has been explicitly fine-tuned for code execution efficiency optimization tasks. The LLM possesses both contextual understanding and instruction-following capabilities tailored to performance improvement. It leverages natural language suggestions to perform structural reorganization or semantically equivalent substitutions on $C_{cleaned}$, to generate a more efficient version $C_{optimized}$. The fine-tuning strategy will be described in detail in Section~\ref{Fine-tuning}.

\textbf{Step 4.3. Post-Processing.} Since LLMs often generate mixed-format outputs - including natural language explanations, markdown formatting, and code snippets - extracting a clean and complete version of the optimized code $C_{optimized}$ is essential for downstream analysis and evaluation. To address this, we design the prompt to follow a structured format (as provided in Step 4.1), instructing the model to enclose the generated code within a Markdown code block (i.e., using triple backticks \verb|```|). In the post-processing phase, we apply a regular expression to extract the contents of the code block. For example, given a model output wrapped in triple backticks (\verb|```|\texttt{python}\verb|```|), we use a regex such as \verb|```|\texttt{(?:python)?}\verb|\|\texttt{s*([}\verb|\|\texttt{s}\verb|\|\texttt{S]*?)}\verb|```| to retrieve only the code. The regex enables efficient retrieval of a well-structured and semantically coherent optimized version.

The optimization module not only integrates optimization suggestions and generates improved code but also effectively controls the risk of hallucinations, making it a critical component in ensuring the final optimization quality.

\section{Experimental Design}\label{Experimental Design}
In this section, we present the design of the experiments used to prepare and conduct a comprehensive evaluation of \fp{}. We first describe the data collection in Section~\ref{Data Collection} and preprocessing processes in Section~\ref{Data Preprocessing}, followed by the construction of the knowledge base that supports optimization tasks in Section~\ref{Knowledge Base Construction}. Next, we detail the preparation of the training and evaluation datasets in Section~\ref{Training Dataset Preparation} and Section~\ref{Evaluation Dataset Preparation}, respectively, ensuring that the data is suitable and consistent for subsequent experiments. Finally, we introduce the experimental setup in Section~\ref{Experimental Setup}, including experimental environment and fine-tuning parameter settings, as well as improvements made to the PIE benchmarking program in Section~\ref{Improved PIE Benchmarking Program}.

\subsection{Data Collection}\label{Data Collection}
To support RAG-enabled optimization, a dataset that contains both code and its associated performance measurements is necessary, as knowledge base retrieval requires performance-aware examples to provide effective optimization guidance. Datasets commonly used for code generation tasks, such as HumanEval~\citep{Chen2021Codex} and MBPP~\citep{Austin2021ProgramSW}, contain only mappings between natural language descriptions and target code, but lack execution performance metrics. As a result, they fall short of meeting the requirements for code execution efficiency optimization. To this end, this work employs data sources derived from online judge websites (e.g., AIZU~\citep{Aizu}, AtCoder~\citep{Atcoder}, and Leetcode~\citep{LeetCode}), which typically record each user submission along with corresponding execution status information (e.g., runtime and resource usage information). Such data provide performance measurements that enable the retrieval of effective optimization examples for improving code execution efficiency.

We select two representative datasets as our primary data sources: the Python split of the PIE dataset~\citep{Madaan2023LearningPC} and the Mercury dataset~\citep{du2024mercury}. The \textbf{PIE dataset} is a benchmark designed for code performance optimization tasks, built upon IBM CodeNet. It contains approximately 35,000 code editing pairs, where each pair consists of two implementations for the same programming problem, with one version (typically the post-edit version) showing significantly better execution efficiency than the other. The Python split of PIE has been widely used in studies on code optimization~\citep{Duan2023PerfRLAS, Madaan2023SelfRefineIR, Gao2024SearchBasedLF, Peng2024PerfCodeGen}. Each record in the dataset includes fields such as the submitted code and its runtime, which makes it potentially suitable for fine-tuning LLMs targeting performance improvement. The key fields of the PIE dataset used in our study are shown in Table~\ref{tab:pie_schema}. 

The \textbf{Mercury dataset} is primarily designed to train and evaluate the ability of LLMs to generate efficient implementations for code generation tasks. It contains a total of 1,889 Python tasks, each associated with multiple user-submitted solutions. The \textit{solutions} field of the dataset is a list of solution objects, where each object includes not only the submitted code (in the \textit{solution} subfield) but also metadata such as execution time and hash values. Other fields in the dataset, such as the original source information of the programming problems, are less relevant to our task and are therefore omitted in subsequent processing. Additionally, the task–submission structure of the Mercury dataset differs slightly from the optimization task addressed in this paper, and the detailed preprocessing steps are described in Section~\ref{Data Preprocessing}. The key fields of the Mercury dataset used in our study are shown in Table~\ref{tab:mercury_schema}.

\begin{table}[h]
    \centering
    \caption{Key Fields of the PIE Dataset (the Python Split)}
    \begin{tabular}{>{\raggedright\arraybackslash}p{1.8cm}>{\raggedright\arraybackslash}p{4.5cm}>{\raggedright\arraybackslash}p{1.1cm}}
    \toprule
         \textbf{Name} & \textbf{Description} & \textbf{Type} \\
    \midrule
         problem\_id & Unique identifier of the programming problem & INT \\
         \hline input & Low-efficiency code implementation for the given problem & STRING \\
         \hline target & Edited high-efficiency code implementation & STRING \\
         \hline cpu\_time\_v0 & Runtime of the input implementation (in milliseconds) & FLOAT \\
         \hline cpu\_time\_v1 & Runtime of the target implementation (in milliseconds) & FLOAT \\
         \hline diff & Difference between input and target, represented in unified diff format & STRING \\
        \bottomrule
    \end{tabular}
    \label{tab:pie_schema}
\end{table}

\begin{table}[h]
    \centering
    \caption{Key Fields of the Mercury Dataset}
    \begin{subtable}[t]{\columnwidth}
        \centering
        \caption{Structure of the Mercury Dataset}
        \begin{tabular}{>{\raggedright\arraybackslash}p{1.8cm}>{\raggedright\arraybackslash}p{4.5cm}>{\raggedright\arraybackslash}p{1.1cm}}
            \toprule \textbf{Name} & \textbf{Description} & \textbf{Type} \\
            \midrule meta\_info & Metadata of the programming task, including problem description, difficulty level, etc. & DICT \\
            \hline input & Multiple user-submitted solutions associated with the task (see the structure in Table~\ref{tab:solutions}) & LIST \\
            \bottomrule
        \end{tabular}
        \end{subtable}
    \par\vspace{1em}
    \begin{subtable}[t]{\columnwidth}
        \centering
        \caption{Structure of the \textit{solutions} Field in the Mercury Dataset}
        \begin{tabular}{>{\raggedright\arraybackslash}p{1.8cm}>{\raggedright\arraybackslash}p{4.5cm}>{\raggedright\arraybackslash}p{1.1cm}}
            \toprule \textbf{Name} & \textbf{Description} & \textbf{Type} \\
            \midrule solution & Submitted source code & STRING \\
            \hline runtime & Execution time of the implementation (in milliseconds) & INT \\
            \bottomrule
        \end{tabular}
        \label{tab:solutions}
    \end{subtable}
    \label{tab:mercury_schema}
\end{table}

It should be noted that these paired code samples are used only during the preparation stage, including optimization knowledge base construction and LoRA fine-tuning. During deployment and inference, the framework requires only the input code to be optimized and does not depend on the availability of corresponding optimized code pairs. 

\subsection{Data Preprocessing}\label{Data Preprocessing}
To support the construction of the training set, test set, and optimization knowledge base for the code execution efficiency task addressed in this study, we build a dataset containing performance-improving code pairs and their performance measurement. Specifically, we preprocess and integrate the PIE and Mercury datasets to derive a combined dataset, referred to as $Optimization-Dataset_{base}$ ($OD_{base}$). We strictly follow the official training/test splits of the PIE dataset, and ensure that the construction based on PIE are performed only on the training split, while the test split is reserved exclusively for evaluation. The standardized structure of $OD_{base}$ contains four core fields: (1) \texttt{input}: low-efficiency code, (2) \texttt{target}: high-efficiency code, (3) \texttt{summary}: a natural language summary describing the optimization from input to target, and (4) \texttt{rate}: the effectiveness score of the optimization.

\begin{figure*}[htbp]
    \centering
    \includegraphics[width=\linewidth]{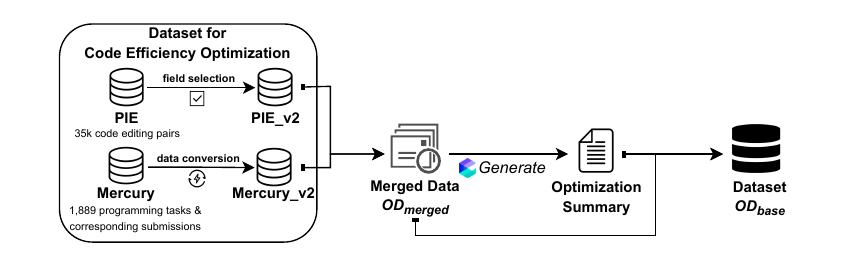}
    \caption{Dataset Construction Workflow}
    \label{fig: Dataset Construction}
\end{figure*}

This section introduces the transformation process from the original datasets - PIE and Mercury - to the target dataset $OD_{base}$, which consists of two main steps: (1) dataset normalization and alignment, and (2) automatic generation of optimization summaries. The overall pipeline is illustrated in Figure~\ref{fig: Dataset Construction}.

\subsubsection{Dataset Normalization and Alignment}
The original structure of the PIE dataset is close to the target dataset $OD_{base}$ used for constructing our optimization knowledge base. Therefore, our preprocessing mainly focuses on field selection and semantic transformation. Specifically:
\begin{itemize}
\item Redundant fields irrelevant to execution efficiency optimization are removed;
\item The following key fields are retained: $input$ (inefficient code), $target$ (optimized code), and $diff$ (unified diff description);
\item A new optimization effectiveness field is constructed as: $rate = cpu\_time\_v0 / cpu\_time\_v1$. This metric measures the runtime improvement of the target implementation relative to the input implementation.
\end{itemize}

The resulting dataset is referred to as $PIE_{v2}$, to distinguish it from the original version.
The Mercury dataset is organized at the task level, where each record contains a programming task and multiple user-submitted solutions. This structure is evidently inconsistent with the pairwise format required by $OD_{base}$. To address this, we convert the data into pairwise samples using the following steps:

\textbf{Step 1.} For each task, extract its corresponding $n$ user submissions.

\textbf{Step 2.} Sort the submissions by execution time in ascending order and designate the fastest submission as $Submission_{fastest}$, which serves as the reference for optimal performance.

\textbf{Step 3.} Pair each of the remaining $n-1$ submissions with $Submission_{fastest}$, resulting in $n-1$ records. Each sample is structured as follows.
\begin{itemize}
\item $input = Submission_i.solution$
\item $target = Submission_{fastest}.solution$
\item $rate = Submission_i.runtime/Submission_{fastest}.runtime$
\end{itemize}

\textbf{Step 4.} Use difflib\footnote{\url{https://docs.python.org/3/library/difflib.html}} to generate a unified-style diff between the input and target, which is recorded as the $diff$ field.

The processed samples form a new dataset $Mercury_{v2}$, whose field structure is fully aligned with that of $PIE_{v2}$ (including $input$, $target$, \textit{diff}, and $rate$). Consequently, the two datasets were combined, resulting in a unified intermediate dataset, $OD_{merged}$, which contains 60k+ samples.

\subsubsection{Automatic Generation of Optimization Summaries}
Since we aim to retrieve code optimization suggestions from the knowledge base, after the normalization and alignment of the dataset are completed, we use an LLM to generate optimization summaries that describe performance improvement. The summary is expected to accurately capture the key modifications made from the inefficient implementation ($input$) to its optimized counterpart ($target$), with a focus on code changes relevant to execution efficiency. Given the scale of our dataset (over 60k+ samples), we employ the Qwen-Max-0125 model~\citep{qwen-max-0125}, which offers a favorable balance between performance and inference cost to generate the optimization summary based on a carefully designed prompt. Qwen-Max-0125 is the most capable model in the Qwen2.5 series, particularly well-suited for complex, multi-step tasks. It demonstrates strong reasoning capabilities and a solid understanding of intricate instructions, achieving promising performance on challenging tasks~\citep{Falcao2025EvaluatingTE}. 
The prompt template is shown below.

\begin{tcolorbox}[colback=gray!10, colframe=gray!80, title=Prompt Template for Generation of Optimization Summaries]

You are an expert in the field of code execution efficiency optimization. Based on the unified-style patch: \{$diff$\}, summarize how it optimizes code execution efficiency. Provide up to two key points. Your output format should be: (If there is one summary) 1. xxx; 2. xxx
 
\label{prompt: summary generation}
\end{tcolorbox}

\rev{To assess the quality of the generated summaries in the knowledge base, we randomly sampled 100 summaries and manually evaluated whether each summary accurately and faithfully described the corresponding code transformation. A summary was considered correct if it accurately captured the main optimization strategy and code transformation without introducing information inconsistent with the corresponding transformation. Among the 100 sampled summaries, 93 were judged correct, yielding a correctness rate of 93\% . This representative assessment provides evidence that the generated summaries generally capture the underlying optimization strategy.}

The LLM-generated natural language summaries are written into the $summary$ field. Finally, by combining $OD_{merged}$ with the generated summaries - while discarding the intermediate $diff$ field — we construct the complete base dataset, $OD_{base}$, which consists of the following four fields: $input$, $target$, $summary$, and $rate$, comprising over 60,000 instances.

\subsection{Knowledge Base Construction}\label{Knowledge Base Construction}
To enable an efficient retrieval of optimization suggestions based on semantic similarity, we build a knowledge base for code execution efficiency optimization using the preprocessed dataset $OD_{base}$ (see Section~\ref{Data Preprocessing}). The construction of the knowledge base is as follows.

First, for each $input$ code snippet in $OD_{base}$, a 768-dimensional vector representation is generated using the code embedding module. The resulting vectors are then appended to the dataset. Next, the augmented $OD_{base}$, which contains both textual and vector representations, is imported into the knowledge base. The mapping of each field to the knowledge base schema is shown in Table~\ref{tab:schema_mapping}.

\begin{table}[h]
    \centering
    \caption{$OD_{base}$ to Knowledge Base Schema Mapping}
    \begin{tabular}{>{\raggedright\arraybackslash}p{1cm}>{\raggedright\arraybackslash}p{1.5cm}>{\raggedright\arraybackslash}p{4cm}}
        \toprule \textbf{$\text{OD}_{base}$ Field} & \textbf{Knowledge Base Field} & \textbf{Description} \\
        \midrule input & - & The inefficient code snippet to be optimized \\
        \hline - & vector & 768-dimensional vector representation obtained from embedding the \textit{input} \\
        \hline summary & summary & Natural language summary of code execution efficiency optimization related to the slow code \\
        \hline rate & rate & Execution efficiency improvement rate \\
        \hline - & id & Unique identifier for each record in the knowledge base \\
        \bottomrule
    \end{tabular}
    \label{tab:schema_mapping}
\end{table}

It is worth noting that, due to the considerations discussed in Section~\ref{Code Generation}, the \textit{diff} and $target$ fields are intentionally excluded from the knowledge base to avoid introducing potential noise or triggering hallucinations. The final constructed knowledge base supports vector-level Approximate Nearest Neighbor (ANN) similarity search, enabling rapid retrieval of the most semantically similar \textit{slow code} given an input code snippet. The corresponding natural language optimization summaries are then returned as reference suggestions for the generation model.

\subsection{Training Dataset Preparation}\label{Training Dataset Preparation}
The objective of the training step is to enhance the performance of the LLM employed in our framework for the final code generation step, aiming at code execution efficiency optimization with reference to natural language suggestions. Specifically, it focuses on the following two aspects:

\begin{itemize}
\item \textbf{Task Adaptation.} Guiding the LLM to understand the contextual prompt structure when optimization suggestions are included in the input, and to learn how to perform performance-oriented code optimization based on natural language instructions.
\item \textbf{Output Standardization.} Improving the consistency of the generated outputs in terms of format and correctness, thereby enhancing the LLM's practicality for optimization tasks.
\end{itemize}

Based on the above objectives, we randomly sampled 4,000 instances from the $OD_{base}$ dataset to construct the training set. We adopt this training size as a practical configuration to provide sufficient task-specific examples for model adaptation while controlling training cost and time. This moderate-sized subset allows the model to learn the prompt structure and optimization patterns without requiring full-dataset fine-tuning. The sampled instances are formatted using the prompt template provided in Section~\ref{Code Generation} and formatting the dataset according to the messages structure\footnote{For reference, see the standard \href{https://huggingface.co/docs/trl/main/en/dataset_formats}{Dataset Formats and Types}.}. The formatting details of the training instances, including the roles and content of each message, are as follows:

\begin{itemize}
\item \textbf{`role':} `system', \textbf{`content’:} System Prompt
\item \textbf{`role':} `user', \textbf{`content’:} User Prompt, where the input code is taken from the $input$ field of $OD_{base}$, and the suggestions are top-$k$ optimization suggestions retrieved from the knowledge base based on semantic similarity and optimization effectiveness. 
\item \textbf{`role':} `assistant', \textbf{`content’:} the optimized version of the input code, i.e., the $target$ field in $OD_{base}$.
\end{itemize}

Furthermore, since both the training set and the knowledge base are derived from $OD_{base}$, the retrieved suggestions may include the original optimization summary $summary_i$ corresponding to the current training sample $input_i$. If not handled properly, it can lead to the model overfitting to that specific suggestion, thereby reducing its ability to generalize to alternative optimization strategies. 

To mitigate the risk of over-reliance on retrieved suggestions during training, improve the model's generalization performance during inference, and reduce the possibility of potential data leakage caused by the trivial reuse of identical optimization pairs, we apply a further filtration step when constructing the training set: \textbf{suggestions with a similarity score of 1 are explicitly removed}. This ensures that the original optimization summary is not reused as training input for the same code snippet, thereby enhancing the model's generalization capability under diverse suggestion conditions.

\subsection{Evaluation Dataset Preparation}\label{Evaluation Dataset Preparation}
We adopt the test dataset from the Python split of the PIE dataset as the basis for model evaluation. It originally contains fields such as $problem\_id$, $slow\_code\_col$ (inefficient code), $reference\_code\_col$ (manual high-performance solution), and $model\_generated\_potentially\_faster\_code\_col$ (model-generated result). Each task also comes with 3 to 4 test cases with expected outputs, as provided in the dataset, for systematic evaluation of correctness and efficiency.

Notably, some instances in the original test set fail to execute correctly in the current runtime environment due to several issues, including missing external dependencies, version incompatibilities, or I/O errors. To ensure the accuracy and reproducibility of the evaluation, we perform a filtering process on the test set. Specifically, we execute both the inefficient code and the reference implementations for each instance under a standard Python environment, and remove any samples exhibiting missing or incompatible packages, runtime errors, or mismatches between the program output and the expected results defined by the test cases. This process is automated using a Python script with multiprocessing to handle all instances efficiently.

After filtering, a total of 752 test instances that can stably run and produce correct outputs under the current runtime environment are retained to form the final test set used for evaluating the model's optimization capabilities.

\subsection{Experimental Setup}\label{Experimental Setup}

\subsubsection{Experimental Environment and Configuration}\label{Experimental Environment}
To ensure reproducibility, all experiments were conducted in a stable environment that was consistent and fixed across all examples and setups. Specifically, experiments were conducted on a server running \texttt{CentOS 7.9.2009 (Core)} (64-bit), equipped with dual AMD EPYC 7543 processors (2.8 GHz, 64 cores in total), 256 GB ECC DDR4 3200 MHz memory, and four NVIDIA Tesla A100 (40 GB NVLink) GPUs. \rev{The GPUs were used for LoRA fine-tuning of the open-source models. After fine-tuning, the other steps do not require GPU acceleration and can be executed on standard CPU-based hardware. Therefore, the four-A100 configuration represents the offline model adaptation cost rather than the deployment requirement of \fp{}.}

\rev{In the knowledge base retrieval step, we retrieve $m=5$ candidate records and the absolute similarity threshold is set to $\theta=0.5$. From the filtered candidates $R_m$, we select $k=2$ optimization suggestions for code generation. During the suggestions filtering process, the similarity tolerance threshold is set to $\Delta_s=0.05$. For code generation, the maximum generation length is set to 1024 tokens.}

\subsubsection{Fine-tuning}\label{Fine-tuning}
We apply Low-Rank Adaptation (LoRA) to fine-tune LLMs using the ms-swift framework~\citep{Zhao2024SWIFT}. LoRA introduces trainable low-rank matrices into specific layers while freezing the original model parameters, thereby significantly reducing the number of trainable parameters. Although multiple reasoning-oriented (e.g., Qwen-Max-0125~\citep{qwen-max-0125}) and code-specialized models (e.g.,GPT-5.1-Codex~\citep{gpt-5.1-codex}, CodeLLaMA-7B-Instruct~\citep{llama}, \textbf{Deepseek Coder-6.7B-Instruct}~\citep{Guo2024DeepSeekCoderWT}, and \textbf{Qwen2.5-Coder-7B-Instruct}~\citep{Hui2024Qwen25CoderTR}) were considered in our evaluation (see Section~\ref{Benchmarks}), due to hardware limitations, only the latter three open-source models were fine-tuned in this study, as their model parameters are available for fine-tuning. Fine-tuning is conducted on the dataset introduced in Section~\ref{Training Dataset Preparation}, with the goal of improving the model's capability to generate optimized code with higher execution efficiency, guided by natural language suggestions.

\begin{figure*}[htbp]   
  \centering            
  \subfloat[Training Loss of LoRA Fine-Tuning on Deepseek-Coder-6.7B-Instruct]   
  {
      \label{fig:train loss of deepseek}\includegraphics[width=0.48\textwidth]{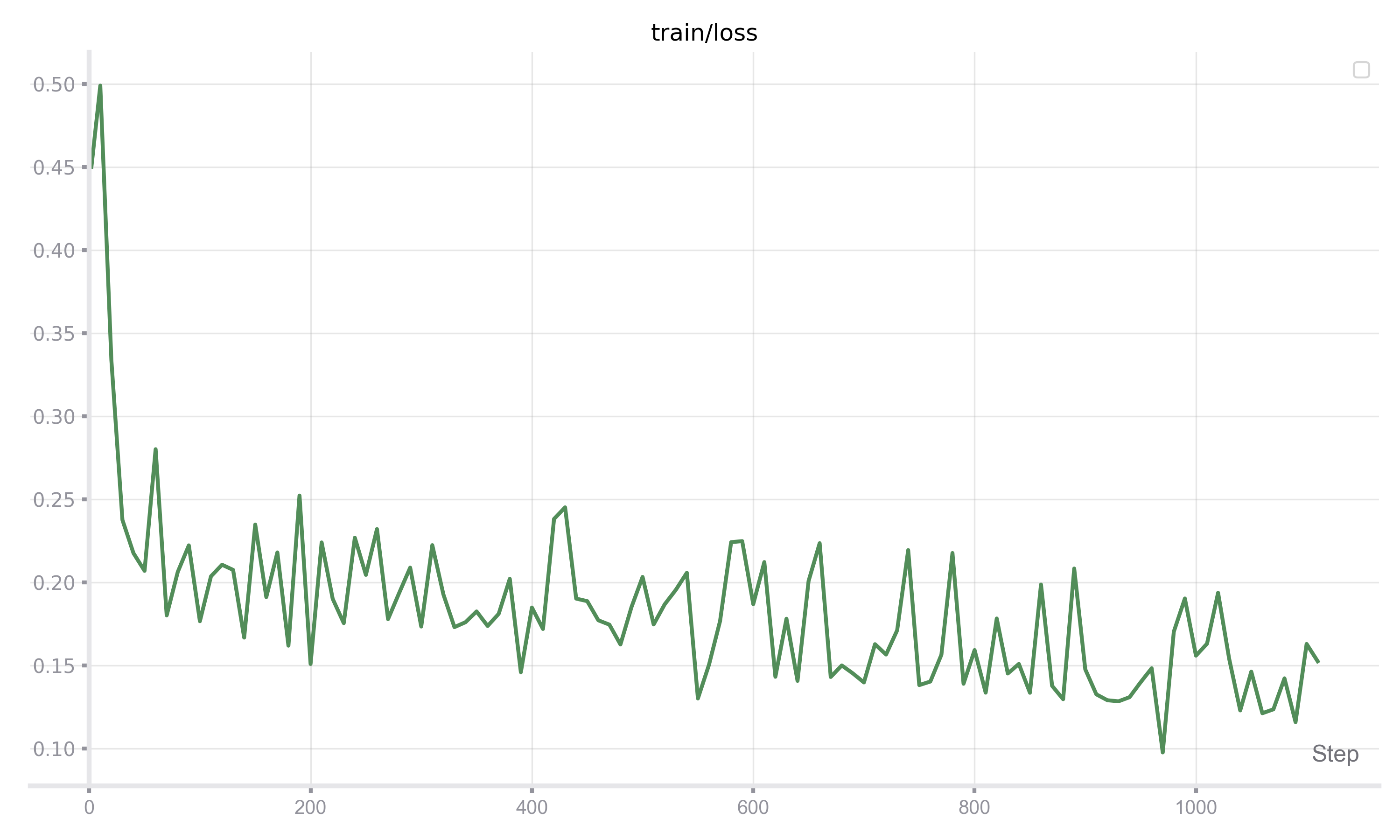}
  }
  \hfill
  \subfloat[Validation Loss of LoRA Fine-Tuning on Deepseek-Coder-6.7B-Instruct]
  {
      \label{fig:val loss of deepseek}\includegraphics[width=0.48\textwidth]{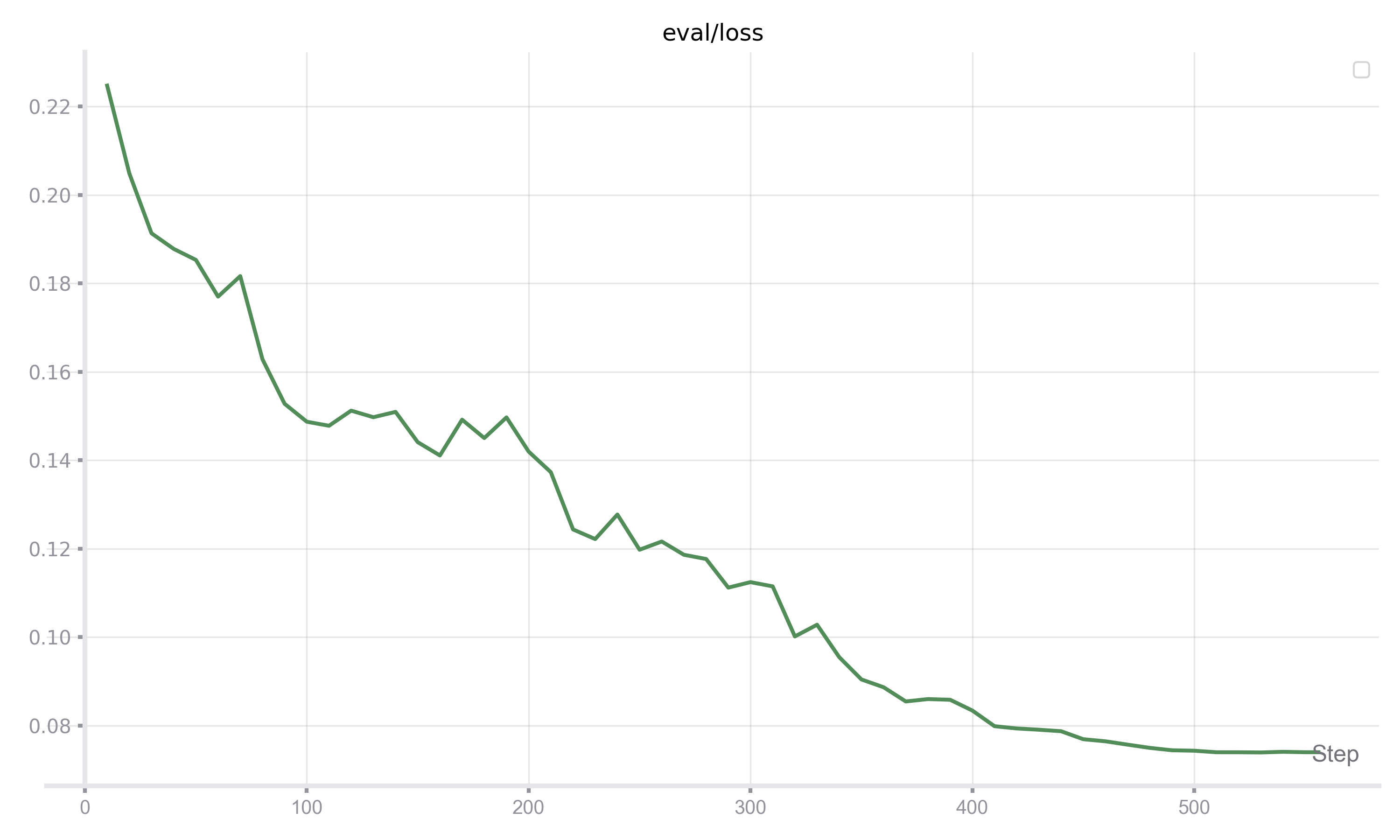}
  }
  \caption{LoRA Fine-Tuning Results on Deepseek-Coder-6.7B-Instruct} 
  \label{fig: fine-tuning results on deepseek}          
\end{figure*}

\begin{figure*}[htbp] 
  \centering           
  \subfloat[Training Loss of LoRA Fine-Tuning on Qwen2.5-Coder-7B-Instruct]   
  {
      \label{fig:train loss of qwen}\includegraphics[width=0.48\textwidth]{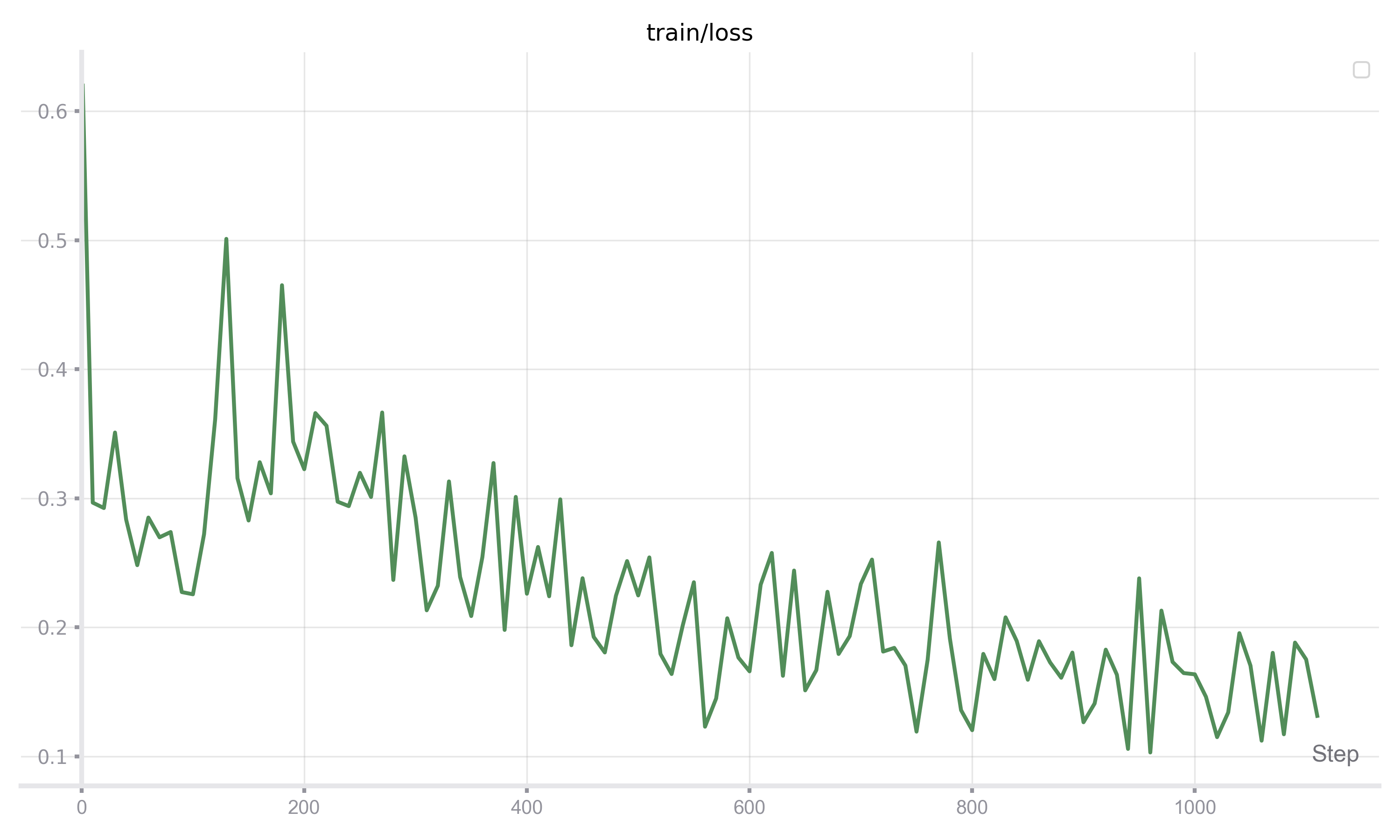}
  }
  \hfill
  \subfloat[Validation Loss of LoRA Fine-Tuning on Qwen2.5-Coder-7B-Instruct]
  {
      \label{fig:val loss of qwen}\includegraphics[width=0.48\textwidth]{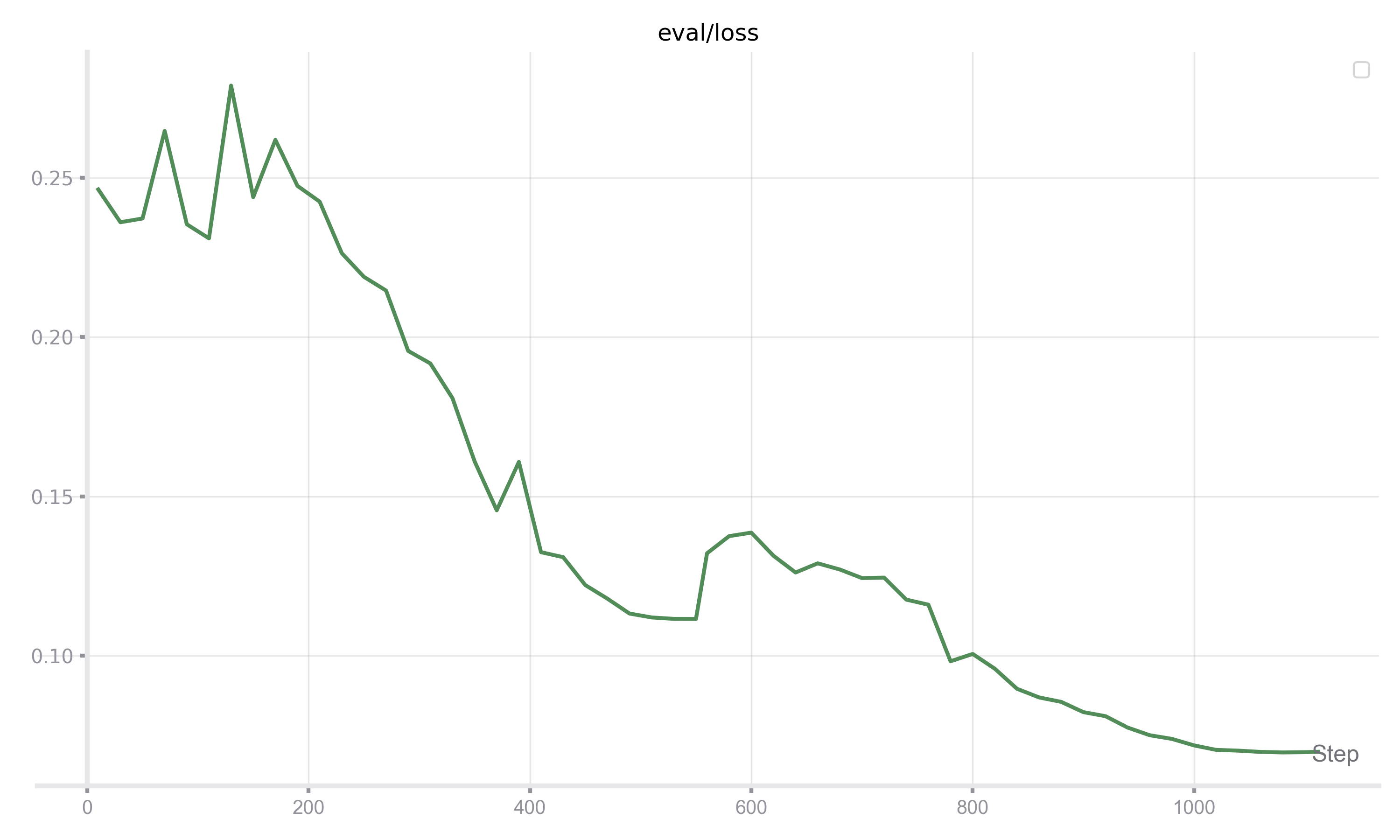}
  }
  \caption{LoRA Fine-Tuning Results on Qwen2.5-Coder-7B-Instruct} 
  \label{fig: fine-tuning results on qwen}          
\end{figure*}

\begin{figure*}[htbp] 
  \centering           
  \subfloat[Training Loss of LoRA Fine-Tuning on CodeLLaMA-7B-Instruct]   
  {
      \label{fig:train loss of codellama}\includegraphics[width=0.48\textwidth]{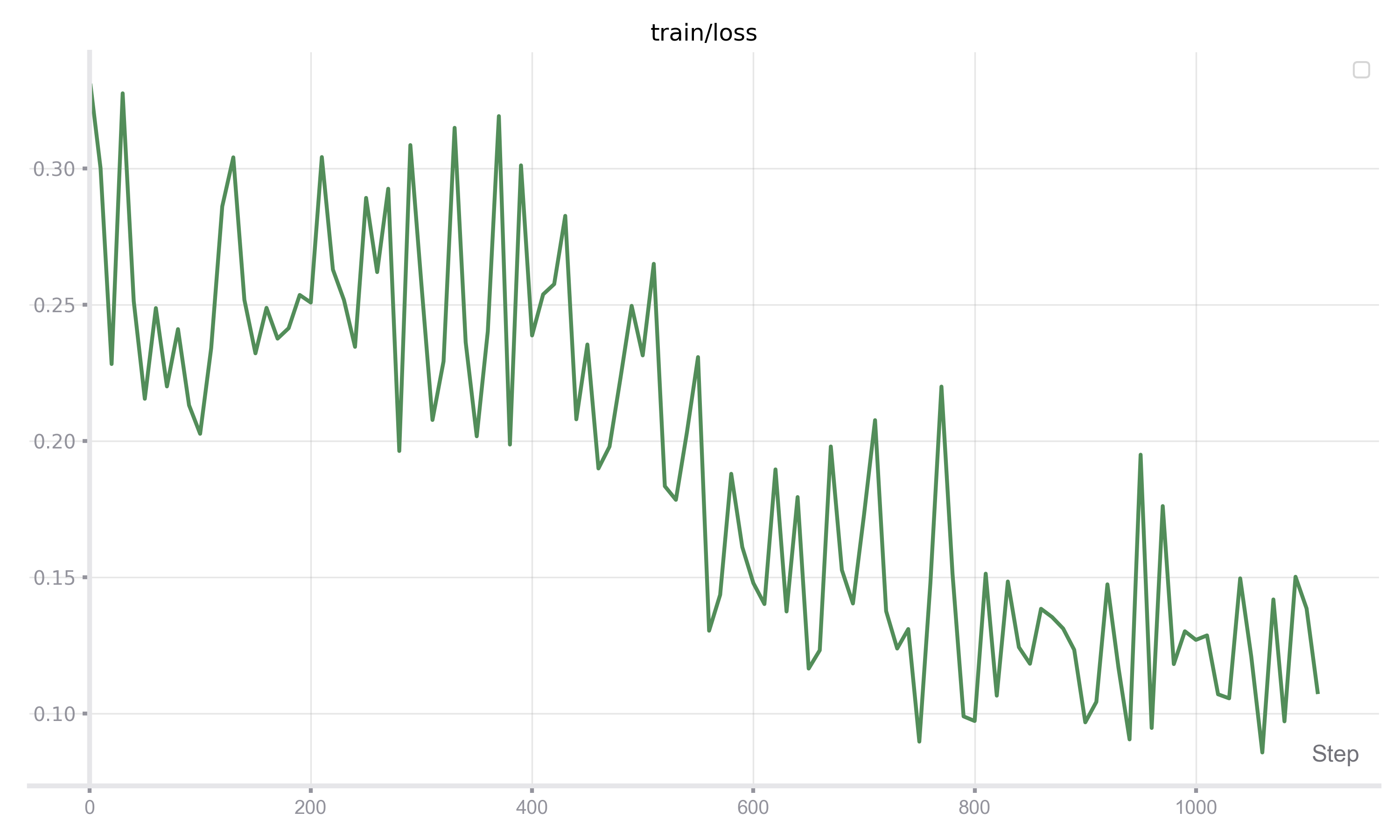}
  }
  \hfill
  \subfloat[Validation Loss of LoRA Fine-Tuning on CodeLLaMA-7B-Instruct]
  {
      \label{fig:val loss of codellama}\includegraphics[width=0.48\textwidth]{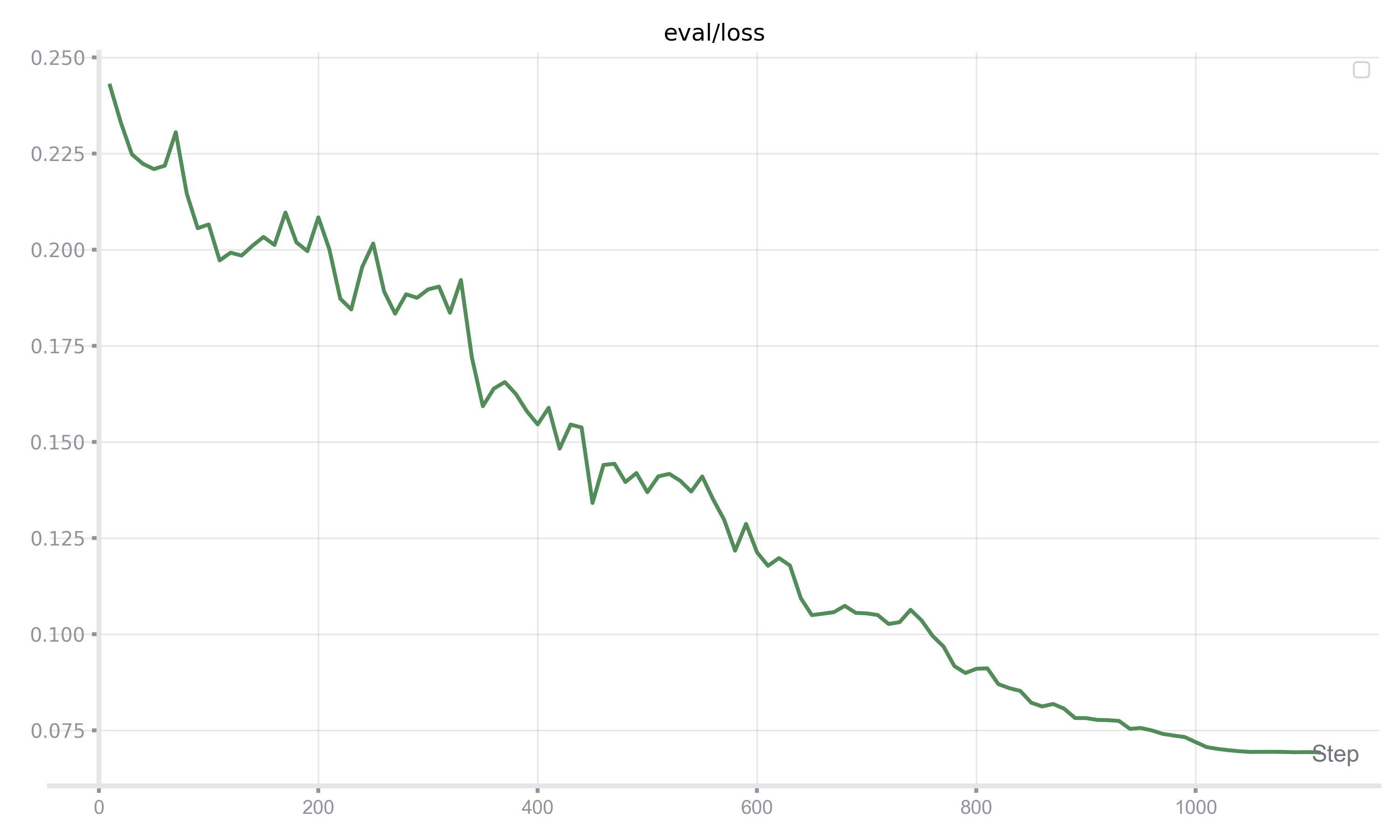}
  }
  \caption{LoRA Fine-Tuning Results on CodeLLaMA-7B-Instruct} 
  \label{fig: fine-tuning results on codellama}          
\end{figure*}

We adopt the ms-swift framework~\citep{ms-swift} provided by the ModelScope community platform to perform LoRA fine-tuning, with the training hyperparameters listed in Table~\ref {tab:parameter settings}. Specifically, the LoRA rank $r$ is set to 8, following prior LoRA studies~\citep{Hu2022LoRA, Kalajdzievski2023ARS}, which show that small rank values can provide a favorable trade-off between adaptation capacity and computational efficiency. The training results are shown in Figure~\ref{fig: fine-tuning results on deepseek} (Deepseek Coder-6.7B), Figure~\ref{fig: fine-tuning results on qwen} (Qwen2.5-7B) and Figure~\ref{fig: fine-tuning results on codellama}, respectively.
\begin{table*}[htbp]
\centering
\caption{Training Hyperparameter Settings for LoRA Fine-Tuning}
\begin{tabular}{ccc}
\toprule 
\textbf{Parameter} & \textbf{Value}	& \textbf{Description} \\
\midrule 
learning\_rate &  1e-4 & The optimizer step size for updating model parameters   \\
batch\_size &  4 & Number of samples per training step  \\
weight\_decay & 0.1 & L2 regularization to prevent overfitting \\
lr\_scheduler\_type & cosine &  Applies cosine decay to adjust the learning rate \\
warmup\_ratio & 0.05 &  Gradually increases the learning rate during early training \\
gradient\_accumulation\_steps & 4 &  Simulates large-batch training under memory limits \\
epochs & 3 & Number of passes over the entire training dataset \\
lora\_rank & 8 &  Controls the size of inserted low-rank matrices \\
lora\_alpha & 32 &  Adjusts the contribution strength of the LoRA adaptation \\
\bottomrule 
\end{tabular}
\label{tab:parameter settings}
\end{table*}

\subsection{Improved PIE Benchmarking Program}\label{Improved PIE Benchmarking Program}
In the PIE benchmarking program provided by Shypula et al. ~\citep{Madaan2023LearningPC}, each code snippet under evaluation is executed in an isolated environment by launching a separate subprocess. The main program is responsible for recording the runtime and output results. This approach offers several advantages, including isolating external context interference, providing a clean execution environment, and supporting the enforcement of timeouts to terminate abnormal programs, thereby ensuring the stability of the testing process. However, this design also has two limitations, as noted below:

\textit{High overhead and low efficiency.} Each code snippet requires spawning a new subprocess for execution. However, creating a subprocess and allocating system resources incurs significant overhead, especially when conducting multiple testing iterations on large-scale datasets. As a result, the overall benchmarking time increases substantially.

\textit{Inaccurate measurement of execution time.} In the original benchmarking program, timestamps are recorded before the subprocess is launched and after it finishes. This measurement includes the overhead of process startup and inter-process communication, leading to an overestimation of the actual execution time of the code. The measurement error is particularly pronounced for code snippets with very short execution time and may even obscure the true performance gains from optimization.

To address the above limitations, we refactor the original PIE benchmarking framework from the two key perspectives, aiming to improve code execution efficiency and enhance the accuracy of runtime measurement.

\subsubsection{Execution Mechanism Based on \texttt{exec()}}
Following prior code evaluation benchmarks such as HumanEval~\citep{Chen2021Codex}, EvalPlus~\citep{Liu2023IsYC}, and CRUXEval~\citep{Gu2024CRUXE}, we adopt an exec()-based dynamic execution mechanism in our evaluation pipeline. Our implementation employs a hybrid design that combines a long-lived control subprocess with dynamic code execution using the \texttt{exec()} function in Python. The implementation consists of the following steps:

\textbf{Step 1.} A persistent control subprocess is created at the beginning of the main program, rather than spawning a new subprocess for each code snippet under test.

\textbf{Step 2.} The control subprocess receives the code snippet and the corresponding test input from the main process via a message queue.

\textbf{Step 3.} Before executing each snippet, the control subprocess redirects standard input to provide the test input, and standard output is redirected to a $StringIO$ object in memory to capture the program output.

\textbf{Step 4.} To prevent the test code from interfering with the control flow, the code is executed within a new thread using \texttt{exec(code)}, inside an isolated namespace. This ensures that even if the code contains termination commands (such as \texttt{sys.exit()}), they will only affect the thread but not the entire testing process.

This design maintains execution isolation while significantly reducing the overhead of repeatedly spawning new processes, thereby improving overall testing efficiency.

In our current evaluation setup, all benchmarked code snippets are single-threaded and do not involve asynchronous or parallel execution. Therefore, the adopted isolated execution mechanism is sufficient for accurate and stable measurement. Handling asynchronous or parallelized code execution will be explored in future work.

\subsubsection{Runtime Instrumentation}
To further improve the accuracy of execution time measurement, we adopt a runtime instrumentation approach. As illustrated in the timing code snippet below, timestamp recording statements are dynamically inserted at the start and end of the test code. Combined with a try-except-finally structure for exception handling, this ensures that execution time is accurately captured regardless of runtime errors.

\begin{tcolorbox}[colback=gray!10, colframe=gray!80, title=Timing Code]
\textbf{import} time

start\_time = time.time()

\textbf{try}:

\quad code under test
    
\textbf{except} Exception as e:

\quad ...

\textbf{finally}:

\quad end\_time=time.time()

\quad duration=end\_time-start\_time

\label{code: timing code}
\end{tcolorbox}

\subsubsection{Evaluation of Benchmarking Effectiveness}
To validate the effectiveness of the improved benchmarking program, we conduct an evaluation focusing on the stability of runtime measurements. We perform the evaluation on the test dataset from the Python split of the PIE dataset mentioned in Section~\ref{Evaluation Dataset Preparation}, which consists of 752 instances. For each instance, we use the reference code as the evaluation target to ensure functional correctness and eliminate confounding factors introduced by incorrect programs.

To assess measurement stability, we adopt the following protocol:

\textbf{1. Randomized Execution Order.}
All test cases are randomly shuffled before each run to mitigate potential bias introduced by execution order (e.g., cache effects).

\textbf{2. Repeated Measurements. }Each benchmarking program executes the full dataset three times independently, producing three runtime measurements for every instance.

\textbf{3. Recorded Metric. }For each instance $i$, we record the measured runtime $t$. This results in three measurements \textit{$t_{i1}$, $t_{i2}$, $t_{i3}$} per instance.

To comprehensively evaluate the effectiveness of the benchmarking programs, we consider both measurement stability and ranking consistency across repeated runs. Specifically, we adopt standard deviation and Kendall’s $\tau$ coefficient~\citep{Kendall1938ANM} as evaluation metrics.

To quantify the stability of the benchmarking program, we adopt the standard deviation of runtime across repeated runs. For each instance $i$, the standard deviation $\sigma_{i}$ is calculated as shown in Equation \ref{eqn: runtime std}, where $t_{ij}$ denotes the runtime measured in the $j$-th run and $\mu_{i}$ is the mean runtime across the three runs. The calculation of $\mu_{i}$ is shown in Equation \ref{eqn: runtime mean}. A lower standard deviation indicates higher stability in runtime measurement.

\begin{equation}
\label{eqn: runtime std}
   \sigma_{i}=\sqrt{\frac{1}{3} \sum_{j=1}^{3}\left(t_{i j}-\mu_{i}\right)^{2}}
\end{equation}

\begin{equation}
\label{eqn: runtime mean}
\mu_{i}=\frac{t_{i 1}+t_{i 2}+t_{i 3}}{3}
\end{equation}

In addition to stability, we further evaluate the consistency of relative performance rankings using Kendall’s $\tau$ coefficient. Given two runs, Kendall’s $\tau$ measures the agreement between the pairwise ordering of instances based on their runtime. Formally, it is calculated as shown in Equation \ref{eqn: Kendall’s Tau}, where $C$ and $D$ denote the number of concordant and discordant pairs, respectively. Since Kendall’s $\tau$ is defined between two rankings, we compute it in a pairwise manner across the three runs. Specifically, we calculate Kendall’s $\tau$ for each pair of runs, i.e., 
($run_1$, $run_2$), ($run_1$, $run_3$), and ($run_2$, $run_3$). The final Kendall’s $\tau$ score is then obtained by averaging over these three pairwise values. A higher value of Kendall’s $\tau$ (closer to 1) indicates stronger agreement between rankings and thus higher reliability of relative performance comparison.
\begin{equation}
\label{eqn: Kendall’s Tau}
    \tau=\frac{C-D}{C+D}
\end{equation}

Table \ref{tab: benchmarking stability and ranking consistency} summarizes the results of the two benchmarking programs. Our improved benchmarking program achieves a significantly lower average runtime standard deviation of 0.0919, compared to the original PIE benchmarking program's 0.1731, indicating improved measurement stability. Furthermore, our improved program obtains an average Kendall’s $\tau$ of 0.9805, while the PIE benchmarking program only achieves -0.0478. This substantial difference shows that the improved program produces highly consistent runtime rankings across repeated runs, whereas the original PIE program yields nearly random rankings.

\begin{table*}[htbp]
\centering
\caption{Comparison of Benchmarking Stability and Ranking Consistency}
\begin{tabular}{cccc}
\toprule 
\textbf{Program} & \textbf{Runtime Std $\sigma_{i}$ (ms)} & $\Delta \sigma_i$ (ms)& \textbf{Kendall's $\tau$} \\
\midrule 
PIE Benchmarking Program &  0.1731& - &  -0.0478 \\
Improved Benchmarking Program & 0.0919 & -0.0812 & 0.9805  \\
\bottomrule 
\end{tabular}
\label{tab: benchmarking stability and ranking consistency}
\end{table*}

Based on the above analysis, we conclude that our improved benchmarking program not only stabilizes runtime measurements but also preserves consistent relative performance ordering across runs, suggesting reliable comparative evaluation.

\section{Evaluation}\label{Evaluation}
According to the experimental design in Section~\ref{Experimental Design}, we conduct a comprehensive evaluation of multiple LLMs on the proposed code execution efficiency optimization task. We first describe the evaluation methodology in Section~\ref{Evaluation Method}. We then introduce the evaluation benchmarks in Section~\ref{Benchmarks}, followed by the evaluation metrics that quantify performance improvements in Section~\ref{Evaluation Metrics}. Finally, we present the evaluation results in Section~\ref{Evaluation Results}, highlighting \fp{}’s performance in code execution optimization.

\subsection{Evaluation Method}\label{Evaluation Method}
To accurately assess the performance of the LLMs in terms of efficiency and correctness under the \fp{} framework, we evaluate three code versions for each problem from the test set mentioned in Section~\ref{Evaluation Dataset Preparation}: $slow\_code\_col$ (inefficient code), $reference\_code\_col$ (manual high-performance solution), and $model\_generated\_potentially\_faster\_code\_col$ (model-generated result), and record runtime and accuracy data. Each code snippet was assigned 3 to 4 test cases ($test\_case_i$) sourced from the PIE dataset. Each test case includes a predefined input ($input_i$) and the corresponding expected output ($output_i$). For every test case, the code was executed four times. To reduce the impact of cold-start effects - such as CPU scheduling, cache loading, and page faults - only the results from the last three executions were recorded, while the first run was excluded.

\textbf{Runtime.} The overall runtime for a code snippet is computed through a two-step process. First, for each individual test case, we determine its stable runtime by averaging the execution time of its last three runs. Then, the recorded runtime is the average of these stable runtimes across all $n$ test cases of the problem (as shown in Equation~\ref{eqn: runtime}), where $n$ denotes the number of test cases, and $Time_{i,j}$ represents the time taken for the $j$-th execution on the $i$-th test case.

\textbf{Accuracy.} We measure functional correctness using a Line-Level Accuracy score, which is based on comparing the execution results of the code against a ground-truth. Specifically, for each test case, this score is calculated as the percentage of generated lines in the code's execution output that exactly match the corresponding lines in the expected execution output. The overall accuracy is the average of these scores across all test cases. The detailed computation is shown in Equation~\ref{eqn: accuracy}, where $n$ denotes the total number of test cases, $L_i$ is the number of output lines in $test\_case_i$, $Output_{i,j}$ is the model's output on the $j$-th line of $test\_case_i$, and $Expected_{i,j}$ is the corresponding expected output line. The indicator function $\mathbb{I}$(·) returns 1 if the lines are a perfect match and 0 otherwise.

\begin{equation}\label{eqn: runtime}
 \text{Runtime} =\frac{1}{n} \sum_{i=1}^{n}\left(\frac{1}{3} \sum_{j=2}^{4} \operatorname{Time}_{i, j}\right)
\end{equation}

\begin{equation}\label{eqn: accuracy}
    \text { Accuracy }=\frac{1}{n} \sum_{i=1}^{n} \frac{1}{L_{i}} \sum_{j=1}^{L_{i}} \mathbb{I}\left(\text { Output }_{i, j}=\text { Expected }_{i, j}\right)
\end{equation}

\subsection{Benchmarks}\label{Benchmarks}
To assess the robustness and general applicability of \fp{}, we benchmark it across LLMs of varying scales, ranging from billions to hundreds of billions of parameters. To assess how effectively \fp{} enhances LLMs' code optimization performance, we evaluate the execution efficiency and functional correctness of the code generated by different LLMs on a unified code optimization dataset. Specifically, the models we select include Deepseek-Coder-6.7B-Instruct~\citep{Guo2024DeepSeekCoderWT}, Qwen2.5-Coder-7B-Instruct~\citep{Hui2024Qwen25CoderTR}, CodeLLaMA-7B-Instruct~\citep{llama}, Qwen-Max-0125~\citep{qwen-max-0125}, and GPT-5.1-Codex~\citep{gpt-5.1-codex}. 
The five representative LLMs covering both open-source and proprietary systems with varying model scales. This selection enables evaluation across different architectures and parameter scales. Table~\ref{tab:models} presents detailed information for each selected model. To ensure a fair comparison, all LLMs are prompted using the prompt template introduced in Section~\ref{Code Generation}.

\begin{table*}[h]
    \centering
    \caption{Five Selected LLMs for Evaluating \fp{}'s Code Optimization Performance}
    \begin{tabular}{cccc}
        \toprule \textbf{Model} & \textbf{Model size} & \textbf{Release date} & \textbf{Open-source}\\
        \midrule Deepseek-Coder-6.7b-Instruct~\citep{Guo2024DeepSeekCoderWT}  & 6.7 billion & 2024-02 & yes \\
         Qwen2.5-Coder-7B-Instruct~\citep{Hui2024Qwen25CoderTR} & 7 billion & 2025-01 & yes \\
         CodeLLaMA-7B-Instruct~\citep{llama} & 7 billion & 2024-04 & yes \\
        \hline Qwen-Max-0125~\citep{qwen-max-0125} & >100 billion & 2025-01 & no \\
         GPT-5.1-Codex~\citep{gpt-5.1-codex} & >200 billion & 2025-08 & no \\
        \bottomrule
    \end{tabular}
    \label{tab:models}
\end{table*}

We apply the full \fp{} framework to Deepseek-Coder-6.7B-Instruct, Qwen2.5-Coder-7B-Instruct and CodeLLaMA-7B-Instruct, including the knowledge base retrieval module (see Section~\ref{Knowledge Base Retrieval}) and LoRA fine-tuning (see Section~\ref{Fine-tuning}). For Qwen-Max-0125 and GPT-5.1-Codex, due to computational constraints, only the knowledge base retrieval module is used in the evaluation. In total, we conduct evaluations and comparisons on the following models and their variants under a unified test set and experimental environment:
\begin{itemize}
\item \textbf{Base models:} Deepseek-Coder-6.7B-Instruct, Qwen2.5-Coder-7B-Instruct, CodeLLaMA-7B-Instruct, Qwen-Max-0125, and GPT-5.1-Codex.
\item \textbf{\fp{}-enhanced variants:} Deepseek-Coder-6.7B-Instruct-\fp{}, Qwen2.5-Coder-7B-Instruct-\fp{}, CodeLLaMA-7B-Instruct-\fp{}, Qwen-Max-0125-\fp{} (no LoRA fine-tuning applied), GPT-5.1-Codex-\fp{} (no LoRA fine-tuning applied).
\end{itemize}

\subsection{Evaluation Metrics}\label{Evaluation Metrics}
To comprehensively evaluate the performance of LLMs on the task of optimizing code execution efficiency, we select four metrics. These metrics assess both runtime performance and functional correctness, and are used to systematically evaluate three types of code: the LLM-generated optimized code ($model\_generated\_potentially\_faster\_code$), the original slow implementation ($slow\_code$), and the human-written efficient implementation ($reference\_code$). The specific metrics are as follows:

\begin{itemize}
\item \textbf{Runtime\_Mean.} This metric measures the average execution time of correctly executed samples, reflecting the overall execution efficiency of the generated code. It is defined as the average runtime across all test samples with an accuracy of 1 (Equation~\ref{eqn: runtime_mean}). \rev{$n$} denotes the number of samples with full functional correctness (i.e., accuracy = 1), and $Runtime_i$ represents the runtime of the $i$-th sample, which is computed as described in Equation~\ref{eqn: runtime}.
\item \textbf{Pass@1.} This metric is a widely used correctness metric that measures the proportion of samples for which the model-generated code passes all test cases in a single attempt, thereby reflecting the model's accuracy under one-shot generation (Equation~\ref{eqn: pass@1}). $N$ denotes the total number of evaluated samples. Our evaluation focuses on the quality and execution efficiency of a single generated optimization result, which more closely aligns with the intended usage scenario of \fp{}. Therefore, we focus on Pass@1 rather than sampling-based metrics such as Pass@$k$ ($k>1$).
\item \textbf{@Speedup.} This metric measures the extent to which the LLM-generated code improves code execution efficiency compared to the original code (Equation~\ref{eqn: speedup}). \rev{$n$} denotes the number of samples with full functional correctness (i.e., accuracy = 1). $\text{Runtime}_{i}^{origin}$ represents the runtime of the original code for the $i$-th sample, and $\text{Runtime}{i}^{model}$ represents the runtime of the code for the $i$-th sample after being optimized by LLMs.

\item \textbf{Percent Optimized [\%Opt]}: The percentage of correctly optimized samples (\%OPT) measures the proportion of functionally correct samples that are more runtime-efficient (at least 10\% faster) than that of the original code (Equation~\ref{eqn: OPT}). $N$ denotes the total number of evaluated samples, and $\mathbb{I}$(·) is the indicator function that returns 1 if the condition is satisfied, and 0 otherwise.
\end{itemize}

\begin{equation}\label{eqn: runtime_mean}
 \text{Runtime\_Mean} =\frac{1}{n} \sum_{i=1}^{n} \operatorname{Runtime}_i
\end{equation}

\begin{equation}\label{eqn: pass@1}
 \text{Pass@1} =\frac{1}{N} \sum_{i=1}^{N} y_i,\text{ } yi=\mathbb{I}\left(\text {Accuracy}_i=1\right)
\end{equation}

\begin{equation}\label{eqn: speedup}
    \text{@Speedup} =\frac{1}{n} \sum_{i=1}^{n} \frac{\text{Runtime}_{i}^{origin}}{\text{Runtime}_{i}^{model}}
\end{equation}

\begin{equation}\label{eqn: OPT}
\begin{aligned}
 \text{\%OPT} = \frac{1}{N}\sum_{i=1}^{N}
\mathbb{I}\Big(&\text{Accuracy}_i=1 \\
&\wedge\ \text{Runtime}_{i}^{model}\times 1.1
<\text{Runtime}_{i}^{origin}\Big)
\end{aligned}
\end{equation}



\rev{Runtime\_Mean and @Speedup are computed only on functionally correct samples,} i.e., test cases with an accuracy of 1. This is because if the code generated fails to pass all test cases, it cannot be regarded as a valid optimization regardless of its shorter runtime. Performance improvements are only meaningful when functional correctness is guaranteed.

\subsection{Evaluation Results}\label{Evaluation Results}
The experimental results are visualized in the heatmap shown in Table~\ref{tab: results}, illustrating the performance of each model across three dimensions: code generation accuracy, execution efficiency improvement, and generalization of optimization capability. In the heatmap, darker colors indicate better performance. Overall, models integrated with the \fp{} framework consistently demonstrate superior performance across all evaluation metrics, significantly outperforming their original versions and other models of comparable scale.

\begin{table*}[htbp]
    \centering
    \caption{Evaluation Results of Code Execution Efficiency Optimization Capabilities. Note: \textit{DS-C-6.7-I} refers to Deepseek-Coder-6.7B-Instruct, \textit{CL-7-I} to CodeLlama-7B-Instruct, \textit{Qw2.5-7-I} to Qwen2.5-Coder-7B-Instruct, \textit{Qw-M} to Qwen-Max-0125, and \textit{Codex} to GPT-5.1-Codex. \textit{-FP} indicates that the model is integrated with the \fp{} framework.}
    \begin{tabular}{ccccc}
    \hline
    \textbf{Model}             & \textbf{Runtime Mean}           & \textbf{\%OPT}                 & \textbf{Pass@1}                & \textbf{@Speedup}              \\    
    \hline
    
    \textbf{input program}     
        & \cellcolor[HTML]{E2E2E2}15.8014 
        & -      
        & -      
        & -      \\
        
    \textbf{reference program} 
        & \cellcolor[HTML]{919191}3.2803  
        & \cellcolor[HTML]{919191}0.8444 
        & \cellcolor[HTML]{919191}1.0000 
        & \cellcolor[HTML]{919191}2.6752 \\    
        
    \hline
    
    \textbf{DS-C-6.7-I}        
        & \cellcolor[HTML]{F5F5F5}22.7498 
        & \cellcolor[HTML]{D9D9D9}0.2221 
        & \cellcolor[HTML]{E2E2E2}0.4029 
        & \cellcolor[HTML]{E2E2E2}1.6405 \\ 
        
    \textbf{DS-C-6.7-I-FP}     
        & \cellcolor[HTML]{BDBDBD}9.2851 
        & \cellcolor[HTML]{D0D0D0}0.3271 
        & \cellcolor[HTML]{C7C7C7}0.7779 
        & \cellcolor[HTML]{D9D9D9}1.7111 \\
        
    \textbf{Qw-2.5-7-I}        
        & \cellcolor[HTML]{D0D0D0}13.1302 
        & \cellcolor[HTML]{E2E2E2}0.3072 
        & \cellcolor[HTML]{D9D9D9}0.5306 
        & \cellcolor[HTML]{D0D0D0}1.7711 \\
        
    \textbf{Qw-2.5-7-I-FP}     
        & \cellcolor[HTML]{C7C7C7}12.2479 
        & \cellcolor[HTML]{C7C7C7}0.3856 
        & \cellcolor[HTML]{D0D0D0}0.7606 
        & \cellcolor[HTML]{BDBDBD}1.8115 \\
        
    \textbf{CL-7-I}            
        & \cellcolor[HTML]{ECECEC}18.0034 
        & \cellcolor[HTML]{F5F5F5}0.2035 
        & \cellcolor[HTML]{D0D0D0}0.6968 
        & \cellcolor[HTML]{ECECEC}1.5418 \\
        
    \rev{\textbf{CL-7-I-FP}}            
        & \cellcolor[HTML]{E2E2E2}15.1710 
        & \cellcolor[HTML]{B4B4B4}0.5026 
        & \cellcolor[HTML]{ABABAB}0.8417 
        & \cellcolor[HTML]{C7C7C7}1.7885 \\   
        
    \hline
    
    \textbf{Qw-M}              
        & \cellcolor[HTML]{B4B4B4}7.0354  
        & \cellcolor[HTML]{BDBDBD}0.5000 
        & \cellcolor[HTML]{C7C7C7}0.7779 
        & \cellcolor[HTML]{B4B4B4}2.0272 \\
        
    \textbf{Qw-M-FP}           
        & \cellcolor[HTML]{ABABAB}6.8918  
        & \cellcolor[HTML]{ABABAB}0.5785 
        & \cellcolor[HTML]{B4B4B4}0.8112 
        & \cellcolor[HTML]{ABABAB}2.1382\\
        
    \textbf{Codex}         
        & \cellcolor[HTML]{A3A3A3}4.4329  
        & \cellcolor[HTML]{A3A3A3}0.6277 
        & \cellcolor[HTML]{A3A3A3}0.8644 
        & \cellcolor[HTML]{A3A3A3}2.2922\\
        
    \textbf{Codex-FP}           
        & \cellcolor[HTML]{9A9A9A}3.3158  
        & \cellcolor[HTML]{9A9A9A}0.7606 
        & \cellcolor[HTML]{9A9A9A}0.9561 
        & \cellcolor[HTML]{9A9A9A}2.6057\\
        
    \hline
    \end{tabular}
    \label{tab: results}
\end{table*}

\subsubsection{Code Generation Accuracy}\label{Code Generation Accuracy}
In terms of code generation \textbf{accuracy}, as shown in Table~\ref{tab: results}, the original billion-parameter models - including Deepseek-Coder-6.7B-Instruct, CodeLLaMA-7B-Instruct, and Qwen2.5-Coder-7B-Instruct - all achieve \textbf{Pass@1} scores below 70\%, indicating their limited performance on the code execution efficiency optimization task. In contrast, after integrating the \fp{} framework, these models exhibit substantial improvements in accuracy. Specifically, Deepseek-Coder-6.7B-Instruct-\fp{} obtains an increase of approximately 37\% in Pass@1 compared to its original version, while Qwen2.5-Coder-7B-Instruct-\fp{} receives a gain of around 23\%.

These results suggest that billion-parameter LLMs may struggle to capture the structure and requirements of code execution efficiency optimization tasks without targeted adaptation. \fp{} significantly enhances their task-specific performance on code execution efficiency optimization by providing retrieved optimization suggestions, thereby improving the correctness of the generated code.

\begin{tcolorbox}[colback=gray!10, colframe=gray!80]
\textbf{Finding 1: } 
\fp{} significantly enhances the task-specific performance of billion-parameter LLMs on code execution efficiency optimization by providing retrieved optimization suggestions, thereby improving the correctness of the generated code.
\label{finding 1}
\end{tcolorbox}

\subsubsection{Optimization Effectiveness}\label{Optimization Effectiveness}
In terms of \textbf{runtime} performance, the code generated by the billion-parameter models is generally comparable to, or even slower than, the original slow code. Specifically, Deepseek-Coder-6.7B-Instruct exhibits an average runtime of approximately 23 ms, while CodeLLaMA-7B-Instruct averages around 20 ms - both significantly slower than the input code, which has an average runtime of 16 ms. Qwen2.5-Coder-7B-Instruct performs slightly better, with an average runtime of about 13 ms. 

After applying the \fp{} framework, the generated code from all models achieves better execution time than the input code, although they still lag behind the manually optimized reference code, which has an average runtime of 3 ms. Among them, Deepseek-Coder-6.7B-Instruct-\fp{} and Qwen2.5-Coder-7B-Instruct-\fp{} both achieve average runtime of less than 13 ms.

In terms of improved code execution efficiency compared to the original code (\textbf{@Speedup}), all original billion-parameter LLMs achieve values below 1.80. This result suggests that current billion-parameter LLMs are unable to effectively perceive and generate performance-optimized implementations. In contrast, all LLMs augmented with the \fp{} framework demonstrate clear acceleration benefits, with @Speedup values exceeding 1.70. Among the billion-parameter LLMs, Qwen2.5-Coder-7B-Instruct-\fp{} achieves the best performance with a Speedup of 1.81, followed by CodeLLaMA-7B-Instruct-\fp{}, which reaches approximately 1.79. The hundreds of billion-parameter models Qwen-Max-0125 and GPT-5.1-Codex, which already exhibit acceleration capabilities in their original versions, still show further improvement after incorporating \fp{}. Specifically, Qwen-Max-0125 improves by about 0.1 to reach approximately 2.14, while GPT-5.1-Codex improves by 0.3 to nearly 2.61.

These findings highlight that billion-parameter models, when not explicitly tailored for the task, exhibit limited sensitivity to code execution efficiency. In contrast, \fp{} provides an efficient solution for improving the adaptability of LLMs to code execution efficiency optimization and enabling practical deployment without reliance on expensive retraining or specialized hardware offering.

\begin{tcolorbox}[colback=gray!10, colframe=gray!80]
\textbf{Finding 2: } 
\fp{} provides an efficient solution for improving the adaptability of LLMs to code execution efficiency optimization and enabling practical deployment without reliance on expensive retraining or specialized hardware offering.
\label{finding 2}
\end{tcolorbox}

\subsubsection{Generalization of Optimization Capability}\label{Generalization of Optimization Capability}
The \textbf{\%OPT} metric measures the proportion of code samples that can be optimized by the model while maintaining functional correctness, thereby reflecting the generalization capability of its optimization ability. As shown in Table~\ref{tab: results}, the original billion-parameter LLMs, including Deepseek-Coder-6.7B-Instruct, CodeLLaMA-7B-Instruct, and Qwen2.5-Coder-7B-Instruct show limited performance on this metric, with only about 20\%\textasciitilde30\% of the samples being successfully optimized. After incorporating the \fp{} framework, their performance improves. In particular, CodeLLaMA-7B-Instruct-\fp{} achieves a \%OPT of nearly 50\%, indicating strong generalization capability in optimization. As for the hundreds of billion-parameter models, Qwen-Max-0125 achieves an improvement of about 10\% compared to its original version, while GPT-5.1-Codex achieves an improvement of about 15\%.

The improvement in \%OPT indicates that \fp{} not only enhances the average optimization effect but also strengthens the model's ability to identify and generalize across different types of performance bottlenecks, demonstrating strong practical adaptability.

\begin{tcolorbox}[colback=gray!10, colframe=gray!80]
\textbf{Finding 3: } 
\fp{} enhances the average optimization effect and strengthens the LLM's ability to identify and generalize across different types of performance bottlenecks, demonstrating strong practical adaptability.
\label{finding 3}
\end{tcolorbox}

In summary, our proposed \fp{} framework achieves significant improvements across all evaluation metrics, particularly demonstrating strong generalization and practicality in terms of improving code execution efficiency across a wide range of test cases.

\subsubsection{Comparison with Related Work}
To further evaluate the effectiveness of \fp{}, we compare it with PIE~\citep{Madaan2023LearningPC} and SBLLM~\citep{Gao2024SearchBasedLF}. PIE is a representative approach that combines retrieval-based few-shot prompting with performance-conditioned generation to improve code efficiency. \rev{SBLLM combines evolutionary search with LLMs through adaptive optimization pattern retrieval and genetic operator-inspired chain-of-thought prompting to iteratively improve code efficiency. PIE and SBLLM are selected as baselines due to their methodological relevance and their compatibility with our experimental setup.}

Although prior methods introduced in Section ~\ref{Related Work} such as MARCO~\citep{Rahman2025MARCOAM}, SUPERSONIC~\citep{Chen2023SupersonicLT}, DeepPERF~\citep{Garg2022DeepPERFAD}, and Artemis AI~\citep{Giavrimis2025Artemis} are related, direct comparison is not feasible due to the limited availability of open-source implementations or their lack of support for Python workloads. We therefore include PIE and SBLLM as closely related and reproducible baselines.

For a fair comparison, PIE and SBLLM are implemented using the same LLMs and evaluated on the same benchmarks as \fp{}. Due to the additional computational overhead introduced by retrieval-based prompting, we conduct the comparison using one representative open-source model (i.e., Deepseek-Coder-6.7B-Instruct~\citep{Guo2024DeepSeekCoderWT}) and one representative closed-source model (i.e., GPT-5.1-Codex~\citep{gpt-5.1-codex}). All configurations are aligned. The comparison results are shown in Table ~\ref{tab: evaluation on pie}. Overall, \fp{} achieves comparable results to PIE and SBLLM across the evaluated metrics, with improvements observed on several metrics. These results demonstrate \fp{}'s effectiveness in guiding the LLMs toward more efficient Python code optimization.

\begin{table*}[htbp]
    \centering
    \caption{Comparison with Related Work. Note: \textit{DS-C-6.7-I} refers to Deepseek-Coder-6.7B-Instruct, and \textit{Codex} to GPT-5.1-Codex. \textit{-FP} indicates that the model is integrated with the \fp{} framework. \textit{-PIE} indicates that the model is integrated with the PIE framework. \textit{-SBLLM} indicates that the model is integrated with the SBLLM framework.}
    \begin{tabular}{ccccc}
    \hline
    \textbf{Model}             & \textbf{Runtime Mean}           & \textbf{\%OPT}                 & \textbf{Pass@1}                & \textbf{@Speedup}              \\    \hline
    \textbf{input program}     & 15.8014 & -      & -      & -      \\
    \textbf{reference program} & 3.2803  & 0.8444 & 1.0000 & 2.6752 \\ \hline
    \textbf{DS-C-6.7-I}        & 22.7498 & 0.2221 & 0.4029 & 1.6405 \\
    \textbf{DS-C-6.7-I-PIE}        & 14.8346 & 0.2760 & 0.6134 & 1.6726 \\
    \rev{\textbf{DS-C-6.7-I-SBLLM}}        & 9.3653 & 0.3152 & 0.6941 & 1.7002 \\
    \textbf{DS-C-6.7-I-FP}     & 9.2851 & 0.3271 & 0.7779 & 1.7111 \\ 
    \hline
    \textbf{Codex}           & 4.4329  & 0.6277 & 0.8644 & 2.2922\\
    \textbf{Codex-PIE}       & 3.8343  & 0.6951 & 0.9142 & 2.4328\\
    \rev{\textbf{Codex-SBLLM}}       & 3.3094  & 0.7566 & 0.9322 & 2.4328\\
    \textbf{Codex-FP}           & 3.3158  & 0.7606 & 0.9561 & 2.6057\\
    \hline
    \end{tabular}
    \label{tab: evaluation on pie}
\end{table*}

\subsubsection{Cost Effectiveness}
To analyze cost-effectiveness, we report the costs incurred by \fp{}, which originate from two stages: data preprocessing and the code optimization workflow.

Before the code optimization workflow, we employ the closed-source LLM Qwen-Max-0125 to generate optimization suggestions during the data preprocessing stage, as described in Section \ref{Data Preprocessing}. This step incurs a total cost of 24.71\$, corresponding to an average cost of \rev{0.0004\$} per sample. Although this introduces an upfront overhead, the constructed dataset can be reused across code optimization tasks. Therefore, the cost can be amortized and becomes negligible in large-scale or repeated usage scenarios.

During the code optimization workflow, we report token consumption and monetary cost measured over the evaluation dataset. For open-source LLMs (Qwen2.5-Coder-7B-Instruct, Deepseek-Coder-6.7B-Instruct and CodeLLaMA-7B-Instruct), we report token usage as a proxy for computational cost, since their monetary cost depends on hardware and deployment configurations. For closed-source LLMs, we report monetary cost based on publicly available API pricing. \rev{Since input and output tokens are typically charged at different rates, monetary cost is calculated based on the corresponding input and output token consumption rather than being directly proportional to the total number of tokens.} Table \ref{tab: cost} presents the average number of tokens and monetary cost per task across different models.

\begin{table*}[htbp]
\centering
\caption{Token Consumption and Monetary Cost during the Code Optimization Workflow. Note: \textit{DS-C-6.7-I} refers to Deepseek-Coder-6.7B-Instruct, \textit{CL-7-I} to CodeLlama-7B-Instruct, \textit{Qw2.5-7-I} to Qwen2.5-Coder-7B-Instruct, \textit{Qw-M} to Qwen-Max-0125,  and \textit{Codex} to GPT-5.1-Codex. \textit{-FP} indicates that the model is integrated with the \fp{} framework.}
\begin{tabular}{ccccccc}
\toprule 
\textbf{Model} & \textbf{Tokens} & \textbf{Cost (USD)}&\textbf{Model} & \textbf{Tokens} & \textbf{Cost (USD)}&\textbf{Token Increase}\\
\midrule 
DS-C-6.7-I & 625 & -&
DS-C-6.7-I-FP & 759 & - & 21.44\% \\ 
Qw-2.5-7-I & 558 & - &
Qw-2.5-7-I-FP & 610 & - &9.32\%\\
CL-7-I & 637 & -&
CL-7-I-FP & 765 & - &20.09\%\\
\hline
Qw-M & 647 & 0.0014 & Qw-M-FP & 713 & 0.0016 &10.20\%\\
Codex & 633 & \rev{0.0038} & Codex-FP & 742 & \rev{0.0039} &17.22\%\\
\bottomrule 
\end{tabular}
\label{tab: cost}
\end{table*}

As shown in Table \ref{tab: cost}, \fp{} slightly increases the total token consumption compared to the benchmarks. Despite a marginal increase in token usage, our method consistently achieves better optimization effectiveness, resulting in improved cost-effectiveness in terms of performance per unit cost.

\rev{To further characterize the computational resources required by the experiments, we report GPU hours and peak GPU memory usage for the three open-source models in Table \ref{tab: fine-tuning cost}. GPU hours measure the aggregate GPU time consumed during fine-tuning and are calculated as the wall-clock time multiplied by the number of GPUs used. Peak GPU memory usage denotes the maximum GPU memory allocated during the training process.} 

\begin{table}
\centering
\caption{Computational Resource Consumption for LoRA Fine-Tuning. Note: \textit{DS-C-6.7-I} refers to Deepseek-Coder-6.7B-Instruct, \textit{CL-7-I} to CodeLlama-7B-Instruct, \textit{Qw2.5-7-I} to Qwen2.5-Coder-7B-Instruct.}
\begin{tabular}{ccc}
\toprule 
\textbf{Model} & \textbf{GPU Hours (h)} & \textbf{Peak GPU Memory (GB)}\\
\midrule 
DS-C-6.7-I & 3.68 & 21.82
 \\ 
Qw-2.5-7-I & 4.20 & 38.15 
\\
CL-7-I & 4.04 & 26.9
\\
\bottomrule 
\end{tabular}
\label{tab: fine-tuning cost}
\end{table}

It is worth emphasizing that the other steps does not require GPU acceleration and can be executed on standard CPU-based hardware. Thus, the four-A100 configuration used for LoRA fine-tuning represents the offline model adaptation cost rather than the hardware requirement during deployment. This separation between offline fine-tuning and online code generation supports the practical deployment of \fp{} in environments without dedicated GPU resources.

Overall, \fp{} achieves a favorable trade-off between cost and optimization performance. Despite a slight increase in token consumption, the improved optimization effectiveness and the amortized dataset construction cost make it a cost-efficient solution for real-world deployment.

\subsubsection{Preliminary Evaluation on C++}
We provide a preliminary evaluation on C++ to investigate whether \fp{} generalizes beyond Python. While our primary focus is Python due to its widespread use and unique optimization challenges in interpreted languages, it is important to understand whether the proposed framework can also benefit compiled languages such as C++.

We conduct experiments on the C++ split of the PIE dataset~\citep{Madaan2023LearningPC}. The data processing procedure and experimental settings are consistent with those described in Section \ref{Experimental Design}. \rev{To construct the C++ optimization knowledge base, we follow the same procedure as that used for Python. From the test set, 500 instances were randomly selected for evaluation.} For evaluation, the generated C++ code was compiled using GNU g++ (GCC) 11.4.1 with the C++17 standard and compiler optimization level \texttt{-O0}, which disables compiler-level optimization during execution. We conduct experiments on part of the LLMs introduced in Section~\ref{Benchmarks} (i.e., Deepseek-Coder-6.7B-Instruct, Qwen2.5-Coder-7B-Instruct, Qwen-Max-0125 and GPT-5.1-Codex). The experimental results in Table \ref{tab: evaluation on cpp} show that \fp{} consistently improves performance over the benchmarks on C++ tasks. Although the magnitude of improvement is smaller compared to Python, the trend remains consistent, indicating that our framework is beneficial across languages.

\begin{table*}[htbp]
    \centering
    \caption{Evaluation Results on C++. Note: \textit{DS-C-6.7-I} refers to Deepseek-Coder-6.7B-Instruct, \textit{CL-7-I} to CodeLlama-7B-Instruct, \textit{Qw2.5-7-I} to Qwen2.5-Coder-7B-Instruct, \textit{Qw-M} to Qwen-Max-0125, and \textit{Codex} to GPT-5.1-Codex. \textit{-FP} indicates that the model is integrated with the \fp{} framework.}
    \begin{tabular}{ccccc}
    \hline
    \textbf{Model}             & \textbf{Runtime Mean}           & \textbf{\%OPT}                 & \textbf{Pass@1}                & \textbf{@Speedup}              \\    \hline
    \textbf{input program}     & 8.4120 & -      & -      & -      \\
    \textbf{reference program} & 2.3342  & 0.4600 & 1.0000 & 2.4812\\    \hline
    \textbf{DS-C-6.7-I}        & 12.8214 & 0.1960 & 0.4420 & 1.6834\\ 
    \textbf{DS-C-6.7-I-FP}     & 6.1351  & 0.2640 & 0.5560 & 1.7103\\
    \textbf{Qw-2.5-7-I}        & 8.8751  & 0.2020 & 0.5480 & 1.7251\\
    \textbf{Qw-2.5-7-I-FP}     & 7.2434  & 0.2580 & 0.5800 & 1.7634\\
    \hline
    \textbf{Qw-M}              & 4.3162  & 0.2620 & 0.6240 & 2.0168\\
    \textbf{Qw-M-FP}           & 3.7883  & 0.3040 & 0.6800 & 2.1026\\
    \textbf{Codex}             & 3.9237  & 0.2960 & 0.6840 & 2.1832\\
    \textbf{Codex-FP}          & 3.1456  & 0.3240 & 0.7360 & 2.2720\\
    \hline
    \end{tabular}
    \label{tab: evaluation on cpp}
\end{table*}

Overall, these results provide preliminary evidence that \fp{} can generalize beyond Python, while broader evaluation on additional languages is left to future work.

\section{Ablation Study}\label{Ablation Study}

\subsection{Component Ablation}\label{Component Ablation}
To better understand the contribution of each component in our framework, we conduct an ablation study by isolating the effects of Retrieval-Augmented Generation (RAG) and Low-Rank Adaptation (LoRA). Due to the fact that closed-source LLMs (i.e., GPT-5.1-Codex and Qwen-Max-0125) do not support parameter fine-tuning, we conduct the ablation study on the open-source models introduced in Section \ref{Benchmarks} (i.e., Deepseek-Coder-6.7B-Instruct, Qwen2.5-Coder-7B-Instruct, and CodeLLaMA-7B-Instruct), which allow controlled modification of model parameters.

We consider the following variants:
\begin{itemize}
\item \textbf{RAG-only:} We enable the knowledge base retrieval module while using the base pretrained model without LoRA fine-tuning.
\item \textbf{LoRA-only:} We apply LoRA fine-tuning to the base pretrained model without incorporating external knowledge retrieval. \rev{Importantly, the LoRA-only variant is separately trained without retrieved optimization suggestions. The prompt follows the same template as that described in Section~\ref{Code Generation}, except that the \textit{Suggestions} section is removed.}
\end{itemize}

We include the base model (without RAG and LoRA) and the full framework (with RAG and LoRA) as references to facilitate a direct comparison with the ablated variants. The reported results are consistent with those in Section \ref{Evaluation Results}. All variants are evaluated under the same experimental settings described in Section \ref{Experimental Setup} to ensure a fair comparison. Table \ref{tab: ablation study} presents the performance of different variants on dataset described in Section \ref{Evaluation Dataset Preparation}.

\begin{table*}[htbp]
\centering
\caption{Ablation Study on Framework Components. Note: \textit{DS-C-6.7-I} refers to Deepseek-Coder-6.7B-Instruct, \textit{CL-7-I} to CodeLlama-7B-Instruct, and \textit{Qw2.5-7-I} to Qwen2.5-Coder-7B-Instruct. \textit{-FP} indicates that the model is integrated with the \fp{} framework.}
\begin{tabular}{ccccc}
\toprule 
\textbf{Model}             & \textbf{Runtime Mean}           & \textbf{\%OPT}                 & \textbf{Pass@1}                & \textbf{@Speedup}              \\    \hline
    \textbf{DS-C-6.7-I}        & 22.7498 & 0.2221 & 0.4029 & 1.6405 \\ 
    \textbf{DS-C-6.7-I-RAG}        & 11.5464 & 0.3244 & 0.5372 & 1.7062\\ 
    \textbf{DS-C-6.7-I-LoRA}        & 14.3650 & 0.2327 & 0.4255 & 1.6504\\ 
    \textbf{DS-C-6.7-I-FP}     & 9.2851 & 0.3271 & 0.7779 & 1.7111 \\ \hline
    \textbf{Qw-2.5-7-I}        & 13.1302 & 0.3072 & 0.5306 & 1.7711 \\
    \textbf{Qw-2.5-7-I-RAG}        & 12.3771 & 0.3364 & 0.5598 & 1.7815 \\
    \textbf{Qw-2.5-7-I-LoRA}        & 12.5171  & 0.3125 & 0.7287 & 1.7806\\
    \textbf{Qw-2.5-7-I-FP}     & 12.2479 &0.3856 &0.7606 & 1.8115 \\ \hline
    \textbf{CL-7-I}            & 18.0034 & 0.2035 & 0.6968 & 1.5418 \\  
    \textbf{CL-7-I-RAG}            & 16.7137 & 0.2673 & 0.7167 & 1.5507 \\  
    \textbf{CL-7-I-LoRA}           & 17.8814 & 0.3949 & 0.7367 & 1.6693 \\  
    \textbf{CL-7-I-FP}            & 15.1710 & 0.5026 & 0.8417 & 1.7885\\  
\bottomrule 
\end{tabular}
\label{tab: ablation study}
\end{table*}

As shown in Table \ref{tab: ablation study}, both RAG and LoRA contribute positively to the overall performance. The full framework consistently outperforms both individual components across all evaluation metrics. This suggests a complementary relationship between retrieval augmentation and fine-tuning. RAG provides high-quality contextual information, while LoRA enables the model to better utilize such information.

Overall, the results demonstrate that both components (RAG and LoRA) are essential, and their combination leads to the best performance.

\subsection{Effect of Retrieved Context Representation}
To investigate the impact of retrieved context representation, we conduct an additional ablation study on how retrieved knowledge is incorporated into the prompt introduced in Section~\ref{Code Generation}. Specifically, we compare the following three settings:

\begin{itemize}
\item \textbf{Code-only:} the prompt contains only the retrieved code snippets;
\item \textbf{Summary-only:} the prompt contains only the natural language summaries generated for the retrieved code snippets;
\item \textbf{Summary + Code:} the prompt contains both the summaries and the corresponding code snippets.
\end{itemize}

For fair comparisons, all other settings are kept identical across experiments. We conduct the study on two representative models, including one open-source model DeepSeek-Coder-6.7B-Instruct~\citep{Guo2024DeepSeekCoderWT} and one closed-source model GPT-5.1-Codex~\citep{gpt-5.1-codex}. The evaluation follows the same metrics described in Section~\ref{Evaluation Metrics}, including Pass@1, @Speedup, Runtime\_Mean, and \%OPT. Table~\ref{tab: ablation study on retrieved context representation} presents the comparison results under different retrieved context representations.

\begin{table*}[htbp]
    \centering
    \caption{Ablation Study on Retrieved Context Representation. Note: \textit{DS-C-6.7-I} refers to Deepseek-Coder-6.7B-Instruct, and \textit{Codex} to GPT-5.1-Codex.}
    \begin{tabular}{cccccc}
    \hline
    \textbf{Model}      &\textbf{Context Representation}       & \textbf{Runtime Mean}           & \textbf{\%OPT}                 & \textbf{Pass@1}                & \textbf{@Speedup}              \\    \hline
        & Summary-only & 9.2851 & 0.3271 & 0.7779 & 1.7111 \\
    \textbf{DS-C-6.7-I}   & Code-only     & 10.2634 & 0.2782 & 0.6032 & 1.6690 \\ 
    &   Summary + Code  & 16.4272 & 0.2414 & 0.5012 & 1.6513 \\ 
    \hline
      & Summary-only   & 3.3158  & 0.7606 & 0.9561 & 2.6057\\
    \textbf{Codex} &  Code-only     & 3.9124  & 0.6831 & 0.9023 & 2.3383\\
    &   Summary + Code          & 3.8256  & 0.7243 & 0.9361 & 2.5072\\
    \hline
    \end{tabular}
    \label{tab: ablation study on retrieved context representation}
\end{table*}

The results show that using natural language summaries alone achieves the best overall performance across most evaluation metrics. In particular, the \textit{Summary-only} setting consistently achieves better Pass@1 and \%OPT scores while obtaining lower Runtime Mean with substantially shorter prompts. Although the \textit{Summary + Code} setting provides additional implementation details, its performance does not surpass the \textit{Summary-only} setting. This suggests that including retrieved code snippets does not provide additional benefits, as natural language summaries alone already provide sufficient optimization guidance.

\subsection{Dataset Ablation}

To evaluate the impact of dataset composition on optimization performance, we conduct an additional ablation study by separately constructing the retrieval knowledge base and fine-tuning dataset using PIE~\citep{Madaan2023LearningPC} and Mercury~\citep{du2024mercury}. Specifically, we compare three settings: (1) PIE-only, (2) Mercury-only, and (3) the combined PIE + Mercury dataset adopted in our main framework.

Since Mercury is collected from multiple users, its code diffs may contain heterogeneous coding styles, variable renaming, and formatting-related modifications. Such diversity may introduce noisy optimization signals compared with the relatively cleaner transformations in PIE. Therefore, this experiment aims to examine whether Mercury can still provide effective optimization knowledge despite the potential noise.

Following the previous ablation studies on the RAG and LoRA components in Section~\ref{Component Ablation}, this experiment focuses solely on the influence of dataset composition while keeping all other configurations unchanged. To provide a representative analysis while maintaining manageable experimental costs, we conduct this study on the open-source model Deepseek-Coder-6.7B-Instruct~\citep{Guo2024DeepSeekCoderWT} and the closed-source model GPT-5.1-Codex~\citep{gpt-5.1-codex}. \rev{For Deepseek-Coder-6.7B-Instruct, changing the dataset composition affects both the optimization knowledge base and the data used for LoRA fine-tuning. For GPT-5.1-Codex, which does not support parameter fine-tuning, the dataset composition only affects the optimization knowledge base. Therefore, this experiment evaluates the overall effect of dataset composition in the corresponding configurations rather than isolating the individual contributions of the knowledge base, fine-tuning dataset, or their interaction.} The experimental results are summarized in Table ~\ref{tab: ablation study on dataset}.

\begin{table*}[htbp]
    \centering
    \caption{Ablation Study on Dataset Composition. Note: \textit{DS-C-6.7-I} refers to Deepseek-Coder-6.7B-Instruct, and \textit{Codex} to GPT-5.1-Codex.}
    \begin{tabular}{cccccc}
    \hline
    \textbf{Model}      &\textbf{Dataset Setting}       & \textbf{Runtime Mean}           & \textbf{\%OPT}                 & \textbf{Pass@1}                & \textbf{@Speedup}              \\    \hline
        & PIE-only & 10.7348 & 0.3191 & 0.5864 & 1.6723 \\
    \textbf{DS-C-6.7-I}   & Mercury-only     & 9.7214 & 0.3125 & 0.5784 & 1.6618 \\ 
    &   PIE + Mercury  & 9.2851 & 0.3271 & 0.7779 & 1.7111 \\ 
    \hline
      & PIE-only   & 4.1704  & 0.6835 & 0.9335 & 2.3280\\
    \textbf{Codex} &  Mercury-only     & 4.0708  & 0.6795 & 0.9229 & 2.3244\\
    &   PIE + Mercury          & 3.3158  & 0.7606 & 0.9561 & 2.6057\\
    \hline
    \end{tabular}
    \label{tab: ablation study on dataset}
\end{table*}

The results show that the Mercury-only setting performs slightly worse than PIE-only across most evaluation metrics, which is expected given the presence of noisy edits and heterogeneous coding styles in Mercury. Nevertheless, Mercury-only still consistently achieves positive optimization gains, indicating that the dataset retains useful optimization patterns despite the noise. \rev{Furthermore, combining PIE and Mercury consistently yields the best overall performance under the current experimental configuration, suggesting that incorporating both datasets can provide more effective optimization knowledge than using either source alone.}
 
These observations demonstrate that the diversity of optimization transformations in Mercury does not significantly degrade optimization quality. Instead, the inclusion of heterogeneous real-world code modifications in Mercury improves the coverage and robustness of optimization suggestions when integrated with PIE.

\section{Discussion}\label{Discussion}

\subsection{Interpretation of Results}\label{interpretationofresults}
The evaluation results presented in Section~\ref{Evaluation Results} clearly demonstrate that the \fp{} framework significantly improves both the correctness and efficiency of the code across multiple base models. Notably, even instruction-tuned models without any explicit training for efficiency optimization exhibit meaningful performance gains after incorporating \fp{}, suggesting that LLMs possess a latent capacity for performance-aware code generation, which can be elicited through well-designed prompts that incorporate retrieved optimization suggestions.

Interestingly, LLMs with billion-parameter sizes (e.g., 6.7B and 7B) show greater relative improvements than hundreds of billion-parameter models, such as Qwen-Max-0125, possibly due to their weaker inherent optimization ability and greater dependence on external guidance. The consistent improvement in \%OPT across all models indicates that \fp{} not only boosts average code execution efficiency but also expands the range of code instances where effective optimization is achieved. This result suggests that retrieving task-relevant optimization knowledge enables better generalization across diverse code inefficiencies.

Compared with traditional approaches such as manually crafted rule-based methods (e.g., ~\citep{Toffola2015Pb, Olivo2015CLARITY}) or algorithmic optimization approaches (e.g., ~\citep{Deng2024CompilerDreamLA, Lamouri2025Pearl}), \fp{} is designed to provide the following two major advantages:

\begin{itemize}
\item \textbf{Lower Cost.} Rule-based methods rely on manual summarization and encoding by domain experts, which are costly to develop and difficult to maintain. Algorithmic optimization approaches, on the other hand, require the LLM to possess a deep understanding of code semantics and efficiency bottlenecks, often depending on hundreds of billion-parameter models with significantly higher training and deployment costs.
\item \textbf{Broader Applicability.} Manually defined rules and fixed algorithms can cover only a limited range of code execution efficiency optimization scenarios, such as simple loop unrolling or constant folding, and often fail to capture more complex or implicit strategies, such as efficient API substitution or execution path restructuring. In contrast, \fp{} learns from historical optimization examples and leverages a retrieval mechanism to generate more diverse optimization suggestions. In addition, \fp{}'s components are modular and extensible, making our framework potentially adaptable to other programming languages. 
\end{itemize}

Our low-cost and flexible approach \fp{} could be potentially suitable for optimizing code execution efficiency in industrial codebases, especially in environments without specialized computational resources. More complex and higher-cost approaches can be introduced incrementally when initial optimization objectives cannot be fully met.

\subsection{Implications}\label{implications}

\subsubsection{Toward Automatic Code Optimization}
The experimental results of our study demonstrate that \fp{}, our proposed framework combining RAG and LoRA, exhibits strong performance and generalizability in automatic code efficiency optimization tasks. This finding not only validates the potential of LLMs in enhancing code efficiency but also suggests a broader reflection on their unique value compared to traditional optimization methods and their future trajectory.

Compared to traditional approaches, LLMs offer several distinct advantages: 1) \textbf{Understanding Code Semantics and Intent}: Unlike traditional methods that rely on formal rules, LLMs, through pre-training on vast code corpora, can learn the underlying semantics of code and the developer's intent~\citep{Nguyen2025Aesc}. This semantic understanding enables LLMs to identify higher-level optimization opportunities that are often imperceptible to traditional methods, such as refactoring complex code logic or substituting entire algorithms with more efficient alternatives. 2) \textbf{Moving Beyond Heuristics}: LLMs are not bound by fixed heuristic rules. Instead, LLMs learn the correlation between code patterns and performance metrics, which allows the LLMs to potentially discover novel and counter-intuitive optimization strategies, such as eliminating redundant intermediate computations. 3) \textbf{Handling Ambiguity and Diversity}: LLMs can process natural language comments, unstructured code snippets, and even incomplete code. This capability allows them to provide optimization suggestions throughout the entire software development lifecycle, not just during the compilation stage.

With the evolution of model architectures, a key future direction is the integration of static and dynamic information in program execution. This involves empowering models to combine static code features (e.g., Abstract Syntax Trees, Control Flow Graphs) with dynamic features (e.g., runtime performance measurements, profiling information) to form a more holistic understanding of program behavior. Furthermore, with the expansion of the context window, LLMs could gain the ability to analyze and refactor entire repositories. This would enable cross-file and cross-module optimizations at the repository level that go far beyond the current scope of single functions or files.

\subsubsection{Leveraging Historical Code Data with LLMs for Code Optimization}
During the construction of the optimization knowledge base and training dataset, this study conducts systematic cleaning and structural processing of the PIE and Mercury datasets. By utilizing the performance-improving code pairs present in the datasets, we summarize code optimization strategies to build the knowledge base. Our approach attempts to leverage Retrieval-Augmented Generation (RAG) to use historical code optimization records to guide future code optimization. The results indicate that historical code optimization records have the potential to guide code optimization efforts. 
Based on this finding, we suggest that code optimization systems could further mine and structurally organize the version evolution histories of projects on open-source platforms such as GitHub to extract performance-improving code edits that can be leveraged for retrieval-augmented code optimization. Additionally, leveraging LLMs to annotate and summarize historical commits can significantly reduce manual labor costs while greatly expanding the coverage of optimization knowledge for code execution efficiency.

Although the optimization suggestions generated by current LLMs may fall short of human-level granularity and accuracy~\citep{Virk2025calibration}, their efficiency and scalability make them a cost-effective solution for industrial applications. In practice, achieving \textit{good enough} automatic optimization for the majority of code is often more valuable than pursuing perfect optimization for a few samples. Therefore, LLM-driven mechanisms for automatic annotation, retrieval, and generation of code optimization knowledge are expected to become core components of future code optimization.

\section{Threats to Validity}\label{Threats to Validity}
The potential threats to the validity of this study are discussed in accordance with the guidelines proposed in \citep{runeson2009guidelines}. To ensure the reliability and reproducibility of the results, this study was carefully designed across multiple stages, including data construction, model training, and evaluation methods. However, we acknowledge that certain factors may still affect the validity of our findings. In this section, we discuss potential threats that may affect the results of this study.

\textbf{Construct Validity:} 
Construct validity concerns whether the evaluation metrics accurately measure the intended capabilities of this study. We primarily adopt metrics such as average runtime (Runtime\_Mean), Pass@1, @Speedup, and \%OPT to assess model performance on the task of code execution efficiency optimization. While these metrics are widely used and representative in existing literature~\citep{Feng2025TowardsBC, Madaan2023LearningPC, Peng2024PerfCodeGen}, they may not fully capture the improvements in real-world scenarios. This study has two threats to construct validity: (1) Execution time is sensitive to external factors such as the runtime environment and cache state. (2) Pass@1 only evaluates the performance of a single generation, without accounting for potential accuracy improvements through retries or reranking.

\textbf{Internal Validity:} Internal validity refers to whether any confounding factors in the experimental design may compromise the validity of causal inferences. In this study, the observed improvements in model performance may be influenced by both the quality of the fine-tuning dataset and the relevance of the retrieved suggestions. While multiple controlled settings and a standardized evaluation pipeline were employed, a slight risk of data leakage remains due to partial sample overlap between the knowledge base and the fine-tuning data. This overlap may impact the model's ability to generalize its understanding of prompt structures beyond the training set. In addition, due to computational constraints, only the open-source LLMs with billions of parameters were fine-tuned in this study, and consequently, we cannot provide a full assessment and comparison between all LLMs. This may limit the fairness and completeness of the comparison across all LLMs, as the full \fp{} framework was not uniformly applied. 

\textbf{External Validity:}
The optimization tasks and evaluation datasets used in this work are primarily constructed from PIE and Mercury, which are derived from competitive programming and open-source problem repositories. While these datasets provide a valuable benchmark, they may not fully capture the characteristics of production-level industrial code; thus, the effectiveness of the proposed \fp{} framework in real-world application scenarios remains to be further examined. \rev{Moreover, this study evaluates only a limited set of mainstream LLMs, and its evaluation focuses exclusively on Python programs. Although \fp{} is designed to be language-agnostic and could potentially be applied to languages such as C++, Rust, and Go, language-specific characteristics may affect the optimization outcomes. Investigating these effects is left for future work.}

Moreover, this study evaluates a limited set of LLMs. The evaluated models include several open-source models as well as stronger proprietary models, including Qwen-Max-0125 and GPT-5.1-Codex; therefore, the evaluation is not limited exclusively to small models. However, the evaluated model set does not comprehensively cover the latest generation of frontier LLMs, whose rapidly evolving capabilities may lead to different optimization outcomes.
Future work should therefore explore a broader range of models, programming languages, and task domains to further validate the framework's general applicability.

\textbf{Reliability:}  The experimental design, including baseline model selection, evaluation metrics and computation methods, and an improved PIE evaluation program, has been provided to ensure the scientific rigor and reproducibility of the evaluation results. While the improved PIE evaluation program enhance the accuracy of runtime measurements, potential sources of variability remain. Specifically, Python’s interpreted execution model and its built-in garbage collection mechanisms may introduce minor fluctuations in timing, particularly for ultra-short programs where such overheads become non-negligible. To improve the reliability and precision of performance measurements, future work may consider adopting lower-level instrumentation techniques, such as instruction-level profiling or hardware-based counters, instead of relying solely on high-level timestamp measurements.

\section{Conclusions}\label{Conclusions}
In this study, we introduce and implement \textbf{\fp{}}, an LLM-based code execution efficiency optimization framework, with the goal of improving the execution efficiency of code.
Existing approaches for code execution efficiency optimization are mostly based on manually crafted rule libraries or iterative algorithms, which are often costly. To address this issue, we propose an automated framework for code execution efficiency optimization driven by LLMs. The framework consists of four steps: code preprocessing, code embedding, knowledge base retrieval, and code generation. In addition, a knowledge base for code execution efficiency optimization was constructed based on the PIE~\citep{Madaan2023LearningPC} and Mercury~\citep{du2024mercury} datasets. Our dataset contains over 60,000 entries, each including slow code, optimized code, a summary of the optimization from slow to fast code, and the embedding vector of the slow code. 
The experimental results show that \fp{} significantly improves both code execution efficiency and correctness, demonstrating its effectiveness and applicability in code efficiency optimization tasks.

In the next step, we plan to: (1) Collect real-world code from open-source platforms such as GitHub to further evaluate and enhance the effectiveness of \fp{} in industrial contexts. (2) Explore the development of code embedding models specifically designed for execution efficiency optimization tasks, or apply rule-based code slicing techniques to more precisely locate performance bottlenecks, thereby improving the relevance and precision of retrieved optimization suggestions. The current retrieval module in \fp{} relies on code semantic similarity using the UniXcoder model, which encodes source code, comments, and abstract syntax trees. While the UniXcoder-based retrieval approach provides a reasonable level of semantic understanding, the model itself is not tailored to the specific requirements of code execution efficiency optimization. For example, two code segments may share the same type of performance issue (e.g., redundant computation within loops) but differ entirely in functionality, leading to low semantic similarity and thus failing to retrieve useful optimization suggestions. (3) Introduce more stable and fine-grained measurement techniques to further improve the evaluation framework. Although we have redesigned the original PIE testing framework to significantly improve testing efficiency and reduce timing errors, runtime measurements may still be affected by system-level factors such as server load, kernel scheduling, and caching. In future work, we will consider incorporating hardware-level performance counters to directly measure low-level metrics such as the number of executed instructions and CPU cycles, which can further enhance the accuracy and reliability of the evaluation results.

\section*{Data Availability}
The replication package, including the dataset used in this work, has been made available at~\citep{replicationPackage}.

\section*{Acknowledgments}
This work has been partially supported by the National Natural Science Foundation of China (NSFC) with Grant No. 92582203 and No. 62402348. The numerical calculations in this paper were performed on the supercomputing system at the Supercomputing Center of Wuhan University.

\printcredits

\bibliographystyle{cas-model2-names}
\bibliography{references}
\balance
\end{sloppypar}
\end{document}